\documentclass[
onecolumn,
amsmath,
preprint,
12pt,
superscriptaddress,
pre]{revtex4-1}

\usepackage{amsfonts}
\usepackage{graphicx}
\usepackage{siunitx}
\DeclareSIUnit\molar{\mole\per\cubic\deci\metre}
\DeclareSIUnit\Molar{\textsc{m}}
\usepackage{hyperref}
\usepackage{color}
\usepackage{wasysym}
\usepackage{afterpage}

\usepackage{siunitx}
\DeclareSIUnit\molar{\mole\per\cubic\deci\metre}
\DeclareSIUnit\Molar{\textsc{m}}
\usepackage[nokeyprefix]{refstyle}
\usepackage{natbib}
\usepackage{multibib}
\newcites{ltex}{Supplementary References}
\usepackage{tocvsec2}

\begin{document}
\title{Kinetic signature of cooperativity in the irreversible collapse
  of a polymer}
% Diffusion limited crumpling of a chromosome during fixation
%%

\author{Vittore~F. Scolari}
\email{vittore.scolari@gmail.com}
%\thanks{}

\author{Guillaume~Mercy}

\author{Romain~Koszul}
%\email{romain.koszul@pasteur.fr}

\affiliation{Spatial Regulation of Genomes, Genomes \& Genetics
  Department, Institut Pasteur, Paris, 75015, France}
\affiliation{UMR3525, Centre National de la Recherche Scientifique,
  Paris, 75015, France}

\author{Annick~Lesne}
\email{Second affiliation: IGMM, University of Montpellier, CNRS,
  Montpellier, France}
\affiliation{Sorbonne
  Universit\'e, CNRS, Laboratoire de Physique Th\'eorique de la Mati\`ere
  Condens\'ee, LPTMC, F-75252 Paris, France}

\author{Julien~Mozziconacci}
\email{mozziconacci@lptmc.jussieu.fr}
\affiliation{Sorbonne
  Universit\'e, CNRS, Laboratoire de Physique Th\'eorique de la Mati\`ere
  Condens\'ee, LPTMC, F-75252 Paris, France}

\date{\today}

\begin{abstract}
We investigate the kinetics of a polymer collapse due to the formation of
irreversible crosslinks between its monomers. 
Using the contact probability $P(s)$ as a scale-dependent order parameter
depending on the chemical distance $s$, our simulations show the emergence of a
cooperative pearling instability. Namely, the polymer undergoes a
sharp conformational
transition to a set of absorbing states characterized by a length
scale $\xi$ corresponding to
the mean pearl size. This length and the transition time depend on
the polymer equilibrium dynamics and
the crosslinking rate. We confirm experimentally 
this transition using a DNA conformation capture experiment in yeast.

\end{abstract}

% 05.70.Ln Irreversible thermodynamics
% 36.20.Ey Molecular dynamics of macromolecules and polymers
% 61.43.Hv Aggregation diffusion-limited

%\keywords{}

\maketitle

The collapse dynamic of a polymer chain has motivated
multiple theoretical and experimental
investigations~\citep{degennes1985kinetics, grosberg1988role,
  chu1995two, buguin1996effondrement, ostrovsky2001cellular, pitard2001glassy,
  dokholyan2002glassy, halperin2000early, bunin2015coalescence,
  majumder2017kinetics}.
% de Gennes seminal work
The seminal work of de Gennes, considering a collapse caused by
solvent quality reduction with no effects of topological constraints,
predicted a continuous conformational transition through successive
crumpling stages
commonly called the ``expanding sausage model''
~\citep{degennes1985kinetics}. 
%This model did not 
%consider the effects of topological constraints in the collapse dynamics. 
% Grosberg fractal globule
%In contrast, 
Grosberg {\it et al.} proposed a two-stage model, where a
fast collapse is followed by a slow unknotting of
topological constraints  through
reptation~\citep{grosberg1988role}. The meta-stable intermediate
state, called ``fractal-globule'', preserves the fractal features of a
coil while being compact as a globule. 
The predicted existence of
meta-stability was  experimentally confirmed by Chu {\it et
  al.}~\citep{chu1995two}.
The stability of the fractal globule has been
% further
 investigated in theoretical
studies, which quantified the relaxation of this state towards an equilibrium globule~\citep{schram2013stability, chertovich2014crumpled}.
%and concluded to the emergence of an unexplainable higher viscosity, possibly
%coming from the process of knots untangling.
% Pearling XXX
As another description of polymer collapse, 
Buguin {\it et al.}
introduced the concept of pearling through the existence of a
characteristic size, there explained by nucleation theory
%, i.e. by  the interplay of bulk and surface effects
~\citep{buguin1996effondrement}. Pearling
has been subsequently studied in different
works~\citep{halperin2000early, pitard2001glassy, ostrovsky2001cellular, dokholyan2002glassy,
%
% Need to spend more words on majumder, ostrovsky
%
  majumder2017kinetics}.
% Kardar volume reduction
More recently Bunin and Kardar proposed an effective model of polymer collapse, 
consisting in a cascading succession of coalescence events of blobs
actively compressed in a central potential~\citep{bunin2015coalescence}.
%in the absence of thermal fluctuations, 
%as an explanation for the polymer collapse~\citep{bunin2015coalescence}.
% Lieberman Aiden fractal globule-again

All these studies investigate the collapse of a polymer under a deep quench: i.e. starting from an equilibrium conformation, 
interactions between the monomers 
%or between the monomers and an external potential 
are abruptly changed and the system 
relaxes to a new equilibrium state. Memory about the collapse process is lost in this final state. In contrast, 
we here study the collapse dynamics of a chain when it is caused by the cumulative effect of irreversible crosslinks between monomers, in the spirit of the pioneering study by~\citet*{lifshitz1976structure}.
In this case, crosslinks cannot be undone and the final state depends on the collapse dynamics. 
This process has important applications in materials science (e.g. vulcanization) and in molecular biology (e.g. cell fixation).

In order to describe the system, we consider here a scale-dependent order parameter: the contact
probability curve $P_t(s)$, defined as the mean
number of crosslinks present at time $t$ between two monomers at a
chemical distance~$s$.
%separated by a linear distance $s$ along the chain.
This order parameter 
has two important advantages:  it 
reflects
the appearance of local structures such as pearls, and it is a
direct observable in the chromosome conformation capture
experiments described at the end of this letter.

%We first describe the {\it in silico} dynamics.
% induced by irreversible crosslinks,  
%with a focus on the combined effects of aggregation and Rouse
%diffusion, measuring the effects on $P(s)$ at different scales.
%
We run a rejection kinetic Monte Carlo
simulation~\citep{cacciuto2006self, scolari2015combined} reproducing
the Rouse
phenomenology on 2048 beads connected initially by a linear chain
of links of maximum length $b$. Each time two non-linked beads
come in close vicinity
(i.e. their distance fall less then $r_{int} = b/64$), a new
link is made with a probability $p$ reflecting the crosslinking rate
% MANUALREF §I.A.2: Reproduction of the equilibrium dynamics
(details in Supplementary Materials (SM) \S I.A).
These links are then treated exactly as the links between consecutive monomers in the chain.

In the absence of crosslinking, the correlations of bead positions
along time and along the chain
satisfy the Rouse scaling relations with
coefficients $C_t$ and $C_s$~\citep{doi1988theory}:

\begin{equation}\label{eq:coeffdef}
 \begin{split}
%C_t (t_0, t)= t^{-1/2} \cdot &
\left\langle \left| \vec R(0, t_0) - \vec R(0, t +
    t_0) \right|^2 \right\rangle & \sim C_t \cdot t^{1/2},
%    \ \mathrm{and},
 \\
%C_s = s^{-1} \cdot &
\left\langle \left| \vec R(s_0, 0) - \vec R(s + s_0,
    0) \right|^2 \right\rangle & \sim  C_s \cdot s.
\end{split}
\end{equation}

%\begin{equation}\label{eq:coeffdeft}
%%C_t (t_0, t)= t^{-1/2} \cdot &
%\left\langle \left| \vec R(0, t_0) - \vec R(0, t +
%    t_0) \right|^2 \right\rangle  \sim C_t \cdot t^{1/2},
%%    \ \mathrm{and},
% \end{equation}
% \begin{equation}\label{eq:coeffdefs}
%%C_s = s^{-1} \cdot &
%\left\langle \left| \vec R(s_0, 0) - \vec R(s + s_0,
%    0) \right|^2 \right\rangle  \sim  C_s \cdot s.
%\end{equation}

After thermal equilibration of the chain, crosslinking is introduced 
%at a time $t_{on}$ 
as a succession of irreversible and
configuration-dependent changes in the chain topology.
As a proxy for steric constraints, we limit the crosslink
events to a maximum number per bead, $N_{max}$, known as the monomer
functionality, and stop the simulation once this number is reached for
all the beads. $N_{max}$ is equal to $4$ in the figures if not
otherwise specified.

\begin{figure}[b] \centering
    \includegraphics[width=\linewidth]{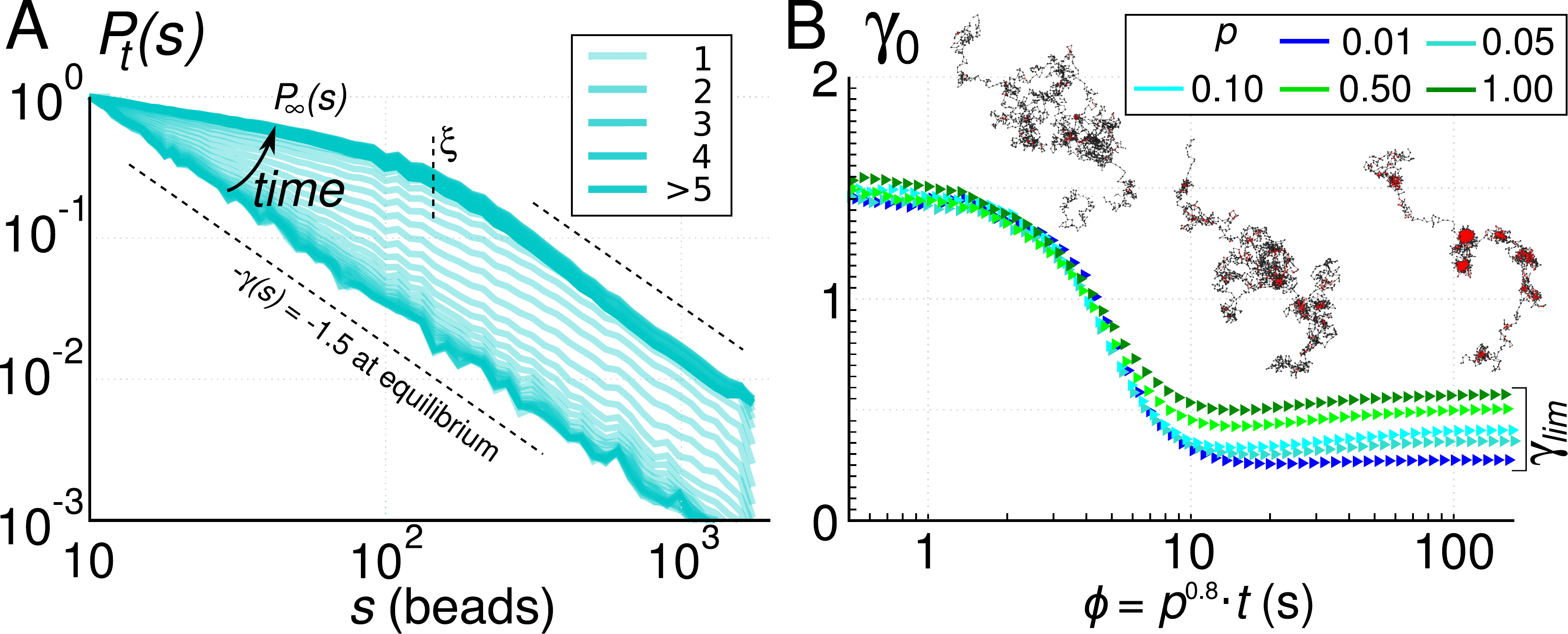}
    \caption{Kinetics of the pearling transition (simulation).
    % seen on the time evolution of the $P(s)$.
      %
      (A) Time evolution of the contact probability curve $P_t(s)$ at
      fixed crosslink
      probability $p = 0.1$, displayed as a superposition of
      semitransparent plots 
      obtained at increasing simulation time~$t$ (black arrow);
      % from crosslinking onset;
      %time ordering is indicated by a black arrow and 
      the resulting color density is
      given in the inset. A crossover at a length $\xi$ arises at
      large enough times.
      Error bars are smaller than the thickness of the line.
      (B) Evolution of  $\gamma_0$, the short-distance slope of the
      log-log plot of the
      $P_t(s)$, as a function of the rescaled
      time variable $\phi$, for  different values of $p$. 
      Inset: snapshots of the time evolution of
      the polymer conformation ($p = 0.1$).
    }
    \label{fig:first}
\end{figure}

\afterpage{%    % defer execution until the next page break occurs anyway
  \begin{figure}[t]
    \centering
    \includegraphics[width=\linewidth]{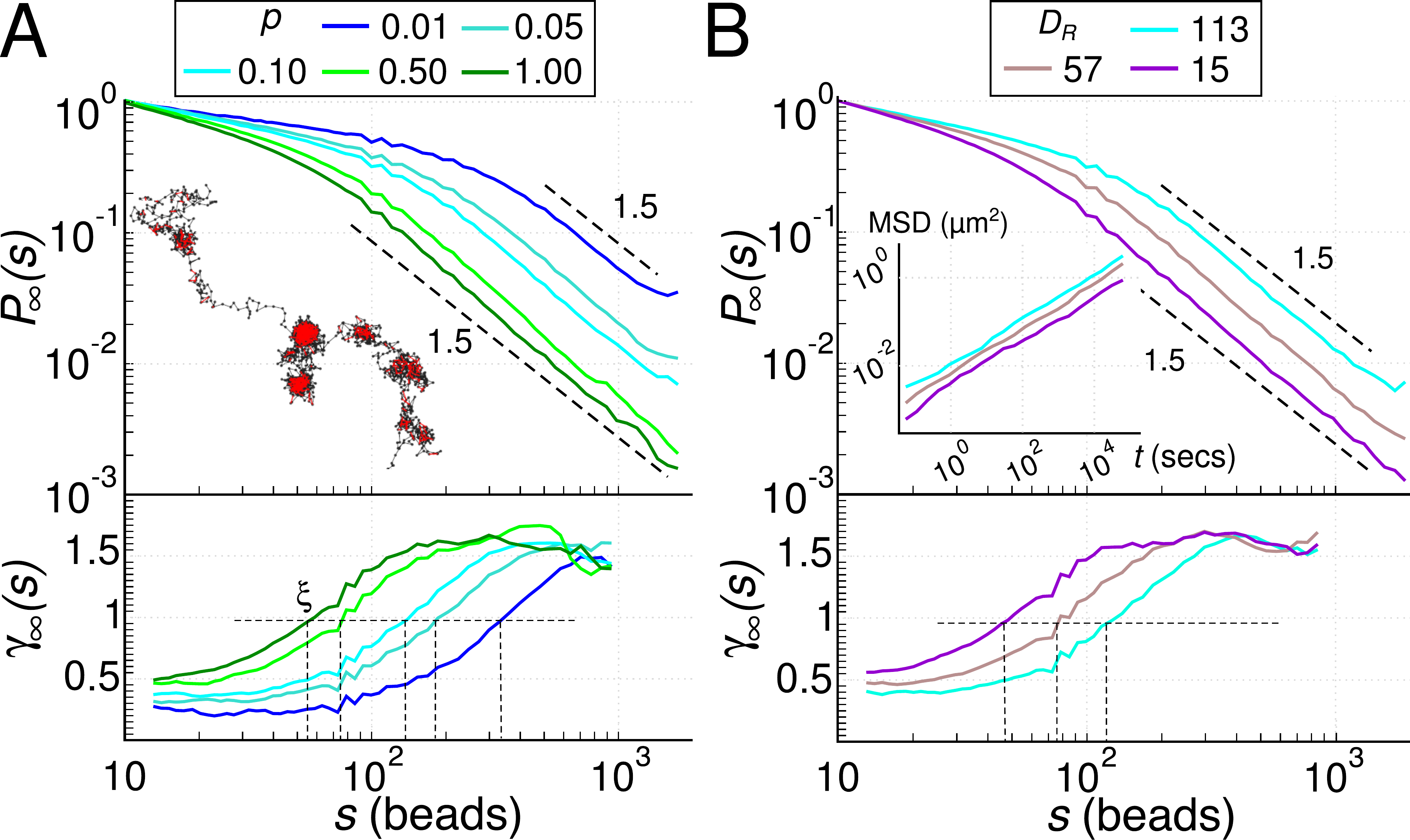}
    \caption{Quantitative features of the pearling transition (simulation).
      % Observations on the mean absorbing state
      % 
      (A) (Upper panel) asymptotic curve $P_{\infty}(s)$  and (Lower panel)
   its local slope $\gamma_{\infty}(s)$ and pearling length $\xi$, for
   different crosslink probabilities $p$. Inset: example of a pearled
   state ($p = 0.1$).
      (B) (Upper panel) asymptotic curve $P_{\infty}(s)$ and (Lower
      panel) its local slope $\gamma_{\infty}(s)$
      for different polymer dynamics,  parameterized by the Rouse
      coefficient $D_R$. Inset:
 monomer mean square displacement (MSD) as a function of  time, whose
 intercept yields a measurement of $C_t$, see Eq.~\ref{eq:coeffdef}.
    }
    \label{fig:sim1}
  \end{figure}
} % end of argument of `\afterpage` command

%Given this irreversible crosslinking dynamics, t
%The expression for the normalized contact numbers between two beads
%separated by a  distance $s$ along the chain,
Given this dynamics, the contact probability
$P_t(s; p, C_s, C_t)$ is a function of $s$, the crosslink probability
$p$, the Rouse coefficients and the elapsed time $t$ from the
crosslinking onset.
%  at $t_{on}$.  
At constant  $p$, the time evolution of this
curve displays a transition from the equilibrium contact probability, 
scaling as  $\propto s^{-\gamma}$ with
$\gamma={3/2}$,~\citep{degennes1980scaling}, to an asymptotic shape
$P_{\infty}(s)$ displaying a crossover between
two different scaling behaviors at short and long chemical distances
~(Fig.~\ref{fig:first}A). This shape and the crossover
length $\xi$ reflect the population average features of the absorbing states reached by the polymer at crosslink saturation.
The exponent $\gamma_0(t)$, corresponding to the value at short distances 
of the local exponent $\gamma(s; t)$ defined from the discrete differential  
\begin{equation}
\gamma(s; t) = - \frac{\Delta \ln [ P(s; t) ]}{ \Delta \ln [ s ] },
\end{equation}
presents a sharp decrease in time (Fig.~\ref{fig:first}B, cyan symbols).

%In order to understand the dependence of 
%To understand how these features depend on the kinetic parameters, we monitored the
We first investigated the effect of the crosslink probability $p$ on the asymptotic curve $P_{\infty}(s)$~(Fig.~\ref{fig:sim1}A, upper panel). 
The crossover length $\xi$ can be estimated as the middle point in the transition of the asymptotic exponent $\gamma_{\infty}(s)$
from short-distance to large-distance values~(Fig.~\ref{fig:sim1}A, lower panel).
This length $\xi$  corresponds to the average length of the polymer segments captured in the pearls, 
and will hereafter be referred as the pearling length.
The characteristic length $\xi$ could also be recovered from the mean squared distance between monomers as a function of the chemical distance $s$ (data not shown).
Individual pearls were identified by clustering together monomers on the contact graph \cite{morlot2016network} using the Louvain algorithm \cite{blondel2008fast}, and their
size was computed in order to confirm that $\xi$ indeed reflects the average number of monomers in pearls.
(see Suppl. Fig.
% MANUALREF
5).
%
%Between the pearls (i.e. f
For $s>\xi$, $\gamma_{\infty}(s) = {3/2}$, consistent with the initial equilibrium state of
the polymer, whereas  $\gamma _{\infty}(s)$ tends inside the pearls to a limiting  value $\gamma_{lim}<1$  at small enough $s$.

  \begin{figure}[t]
    \centering
    \includegraphics[width=0.6\linewidth]{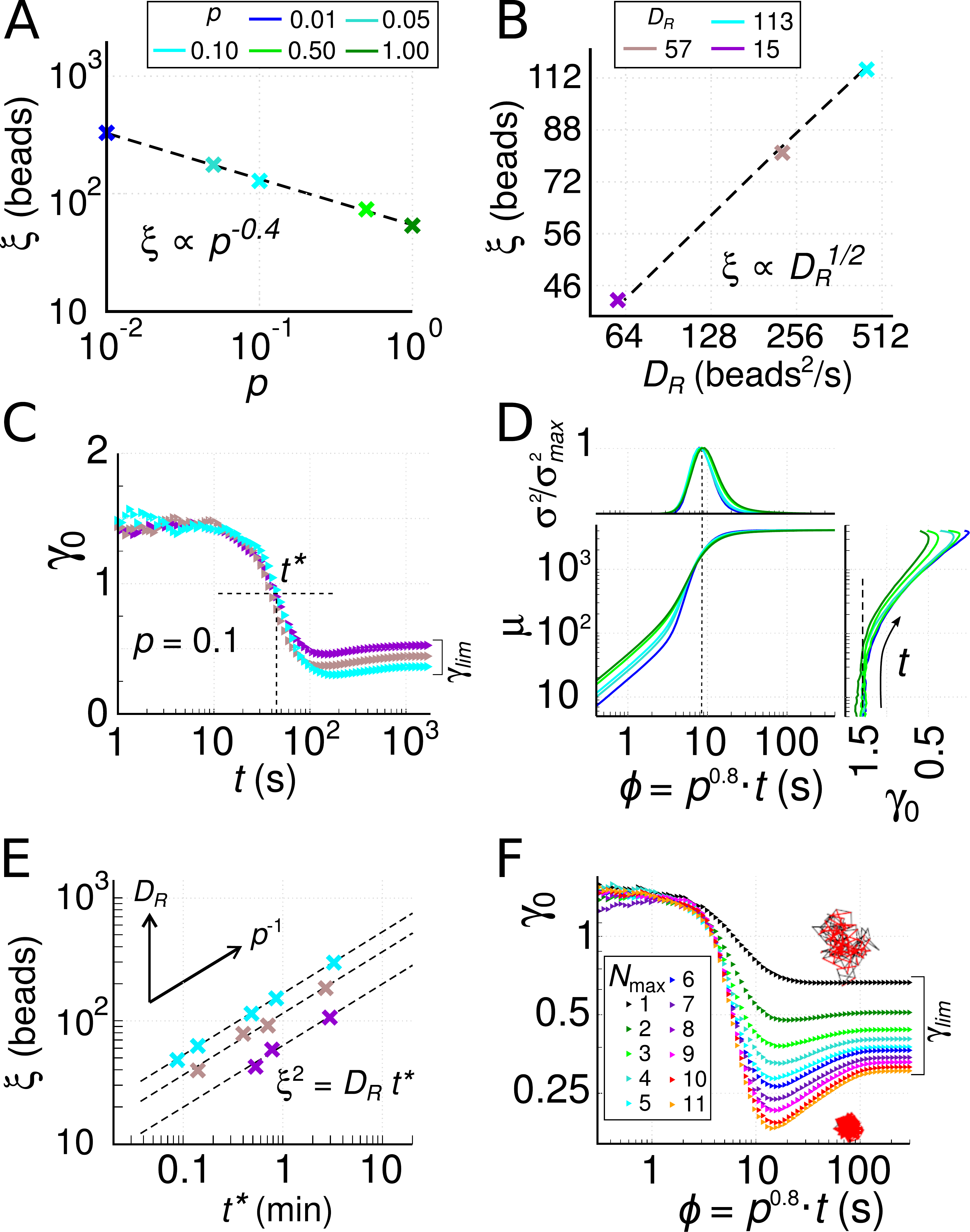}
    \caption{Dependence of the transition dynamics  on
      the kinetic parameters (simulation)
      (A) Variation of the pearling length $\xi$ with the crosslink
      probability $p$.
      (B) Variation of the pearling length $\xi$ with the Rouse
      coefficient $D_R$.
      (C) Time evolution of $\gamma_0$ at different $D_R$, Eq.~\ref{eq:relaxexact} at fixed $p = 0.1$.
      (D) (Lower panel) Mean cumulative number $\mu$ of crosslink events 
      and (Upper panel) its normalized variance 
      $\sigma^2/\sigma^2_{max}$ as a function of $\phi$, and (Right panel)
      scatterplot of $\mu$ and $\gamma_0$.
      (E) Scatter plot of the pearling length $\xi$ and the transition
      time $t^*$; dashed lines are plotted using Eq.~\ref{eq:xit}.
      (F) Evolution of $\gamma_0$  as a function
      of $\phi$ for different values of the monomer functionality $N_{max}$.
    }
    \label{fig:dyn}
  \end{figure}

% 
%Interestingly, t
The length $\xi$  scales with the crosslink probability $p$ 
%is dependent from $p$ and exhibits a
as $\xi \propto p^{-\delta}$, with $\delta = 0.4$ (Fig.~\ref{fig:dyn}A), indicating that the
extent along the chain of the crosslink-induced collapse  is paradoxically more
prominent for small $p$, i.e. low crosslinking rate.
%
% Theoretical argument
%
%We explain this seemingly paradoxical behavior as follows: 
Indeed, conformation changes of polymer loops of
size greater than $\xi$ are diffusion-limited, while for smaller loops, Rouse diffusion is faster than the crosslinking reaction.
In this latter reaction-limited regime,  many conformational fluctuations and contacts can occur and be fixed by crosslinks, producing 
pearls of mean size $\xi$.
% that depends on $p$.
%To get a compatible dependence of $\xi$ in $p$, w
Based on this qualitative picture, we propose a mean-field calculation of the dependence of $\xi$ in $p$.
The relaxation time for a fixed loop of size $s$ scales as:
\begin{equation}
\tau_R(s) = D_R^{-1} \cdot s^2,\ \ \ \mathrm{with}\ \ \
D_R = \frac{\pi^3}{4}\left(\frac{C_t}{C_s}\right)^2,
\label{eq:relaxexact}
\end{equation}
% MANUALREF
(detailed derivation in SM \S II.E.4)
while the average duration $\tau_{cross}$ needed to crosslink contacting
beads is inversely proportional to the crosslink probability:
\begin{equation}
  \tau_{cross} \propto p^{-1}.
\label{eq:tauxl}
\end{equation}
Writing that the pearling length $\xi$ emerges from the
competition between these two dynamical processes yields:
\begin{equation}
  \xi(p) \propto p^{-\delta},
  \label{eq:xip}
\end{equation}
with $\delta = 1/2$ correctly recapitulating
the decrease of $\xi$ at increasing $p$.
We here assumed
that the dynamics is  consistent with Rouse diffusion during the pearling
formation and collapse.  However, Rouse diffusion is not expected to apply to the mesh into what the initially linear polymer is transformed after enough crosslinks, which may explain the different value $\delta = 0.4$
measured in the simulations (Fig.~\ref{fig:dyn}A). 
%originates in the  different topology of the crosslinked polymer, which does not follow excatly 
%instead to the the relaxation time of a compact three-dimensional
%fractal \cite{degennes1976dynamics, dolgushev2015contact}.
%
% Effect of dynamics
%
With the same argument we also predict that  $\xi$ varies
with the dynamical properties of the polymer.
Simulations actually show that variation of the Rouse diffusion coefficient $D_R$
has a dramatic effect on $\xi$ (Fig.~\ref{fig:sim1}B).
For small $D_R$, $\xi$ is small and crosslinking has mostly a local effect.
% most of the crosslink effect occurs locally.
%
When $D_R$ increases, longer polymer segments can reach their
equilibrium conformation between two crosslink events so that $\xi$ becomes larger.
In the line of the above calculation, we expect a scaling
\begin{equation}
  \xi(D_R) \propto D_R^{1/2},
  \label{eq:xiDR}
\end{equation}
which is well reproduced in the simulations
(Fig.~\ref{fig:dyn}B).
%

%
% gamma s in function of time
%

%As regards the kinetics, o
Our simulation moreover
%enables us to investigate in more details the kinetics of the collapse and
 shows that the collapse happens abruptly.
The short-distance exponent $\gamma_0$ presents
a sharp decrease at a time $t^*$, which we call the pearling time.
Before this transition ($t \ll t^*$),
$\gamma_0$ coincides with the exponent at long distances,
${3/2}$, as expected for an equilibrium state.
Only after the transition a smaller exponent  is observed, with a
limiting value  $\gamma_{lim} < 1$ depending on the kinetic parameters.
$t^*$ depends on the crosslink
probability with a scaling $t^*\propto p^{-0.8}$
prompting to define a re-scaled variable $\phi = p^{0.8} \cdot t$.
The evolution  of $\gamma_0$  as a function of $\phi$  
re-scales at any $p$ into a single
transition curve (Fig.~\ref{fig:first}B).
The scaling of
% the transition time
 $t^*$  can also  be explained with the above mean-field argument: 
as $t^*$ emerges from pearling (see polymer snapshots along the transition
curve in Fig.~\ref{fig:first}B), it is equal to the relaxation
time of pearls of mean size $\xi$: $t^* = \tau_R(\xi)$.
From Eq.~\ref{eq:relaxexact},
\begin{equation}
  t^* \propto p^{-2\delta},
\label{eq:tstar}
\end{equation}
and $\phi^* = p^{2\delta} \cdot t$.
As predicted by the above argument and confirmed in the simulation, the transition time
does not depend on the Rouse diffusion coefficient $D_R$ (Fig.
\ref{fig:dyn}C).
% XXX
The pearling transition is the result of the cooperative
effect of multiple crosslinks, that takes place only after
relaxation of loops with length $s < \xi$. This effect is highlighted
in Fig.~\ref{fig:dyn}D, lower panel, that shows the acceleration of
crosslink events at the transition.
This process is accompanied by the decrease of $\gamma_0$
(\ref{fig:dyn}D, right panel) and  a large increase of crosslink number
variability, due to the fluctuation in the size and time of pearl formation
 and consistent with a phase transition
(\ref{fig:dyn}D, upper panel).
Collecting the results from simulations performed at various values
of crosslink probability $p$ and Rouse diffusion coefficient
$D_R$, the  transition points in the plane defined by pearling
time $t^*$ and pearling length $\xi$ (Fig.~\ref{fig:dyn}E) satisfy
the Rouse scaling relation:
\begin{equation}
t^* = D_R^{-1} \cdot \xi^2;
\label{eq:xit}
\end{equation}
that fully recapitulates the relationship between these physical quantities.
% the pearling time, the pearling length and the polymer dynamics.
We finally determine the influence of steric constraints on the final
state by changing the monomer functionality $N_{max}$.
%maximum number of crosslinks, $N_{max}$, allowed for each bead.
%
While $\xi$ and $t^*$ do not depend on $N_{max}$,  the pearl formation and  final internal
conformation do,  as shown by the time behavior of   $\gamma_0$.
After a transition in $t^*$, this short-distance exponent transiently goes toward $0$  for large enough values of $N_{max}$
before plateauing to an asymptotic value $\gamma_{lim}$  varying from $0.3$ to $0.7$ when  $N_{max}$ varies
(see Fig.~\ref{fig:dyn}F and
Suppl. Fig.
% MANUALREF
6).
Examination of the conformational trajectories shows that this behavior can be explained by a two-stage dynamics taking place after the transition in $t^*$. 
The first stage is the formation of densely connected pearls (in red on the snapshots of Fig.~\ref{fig:first}A) 
linked by stretched linkers containing fewer monomers. In these pearls, virtually any monomer can contact any other monomer and $\gamma_0$ strongly decreases. 
A slower process then kicks in: the diffusion-limited crumpling of the stretched linkers between adjacent pearls
(see the  snapshots in Fig.~\ref{fig:dyn}F). 
%Since in the stretched$
In the stretched linkers, mostly adjacent monomers are able to come into proximity, 
hence the contribution of this collapse to $P(s)$ is such that $\gamma_0$ mildly increases.

%To recapitulate our simulation results, 
In summary, our simulation has shown how the interplay between the polymer
Rouse dynamics and the rate at which crosslinks are made induces
a cooperative phase transition to pearled conformations with characteristic scale
 $\xi$. 
We thus obtained a two-stage pearling kinetics, what has already been described in the literature, however
with some significant differences in the underlying mechanisms.
% in the interpretation of results and the underlying mechanisms.
%and the nature of interactions.
%
Our irreversible scenario is not compatible with a simple nucleation and growth process: 
in the nucleation-inspired  model of  Buguin {\it et
  al.}~\citep{buguin1996effondrement}
%  provided a model for the pearling effect inspired by nucleation and as such, 
pearls created with a minimal size of $\xi$ grow continuously
until the overall polymer collapse. 
We can also exclude knotting effects: Grosberg {\it et
  al.}~\citep{grosberg1988role} focused on the role of
knots in the conformational relaxation and predicted a dense globule with a fractal
dimension of 3 and a relaxation through reptation.
In contrast, we neglect volume interactions which are a necessary
element for knot stability. 
To see whether the appearance of a specific length scale depends on the fact that 
we used phantom chain, we performed to extra simulation taking explicitly into account steric effect.
We found in this case that the pearling dynamics of the transition is unchanged (see Suppl. Fig.
% MANUALREF
7).
We also recovered the local formation of a crumple globule-like state in each pearl with $\gamma_0=1$.
The emergence of the characteristic length  $\xi$
however excludes fractality of the absorbing conformations.
% as in the fractal-globule original description.
%
The scale-dependent behavior observed in our simulation reflects
the presence of two different dynamics:
reaction-limited pearling at short distances along the chain,
diffusion-limited collapse at large distances.

\begin{figure}[t]
    \centering
    \includegraphics[width=\linewidth]{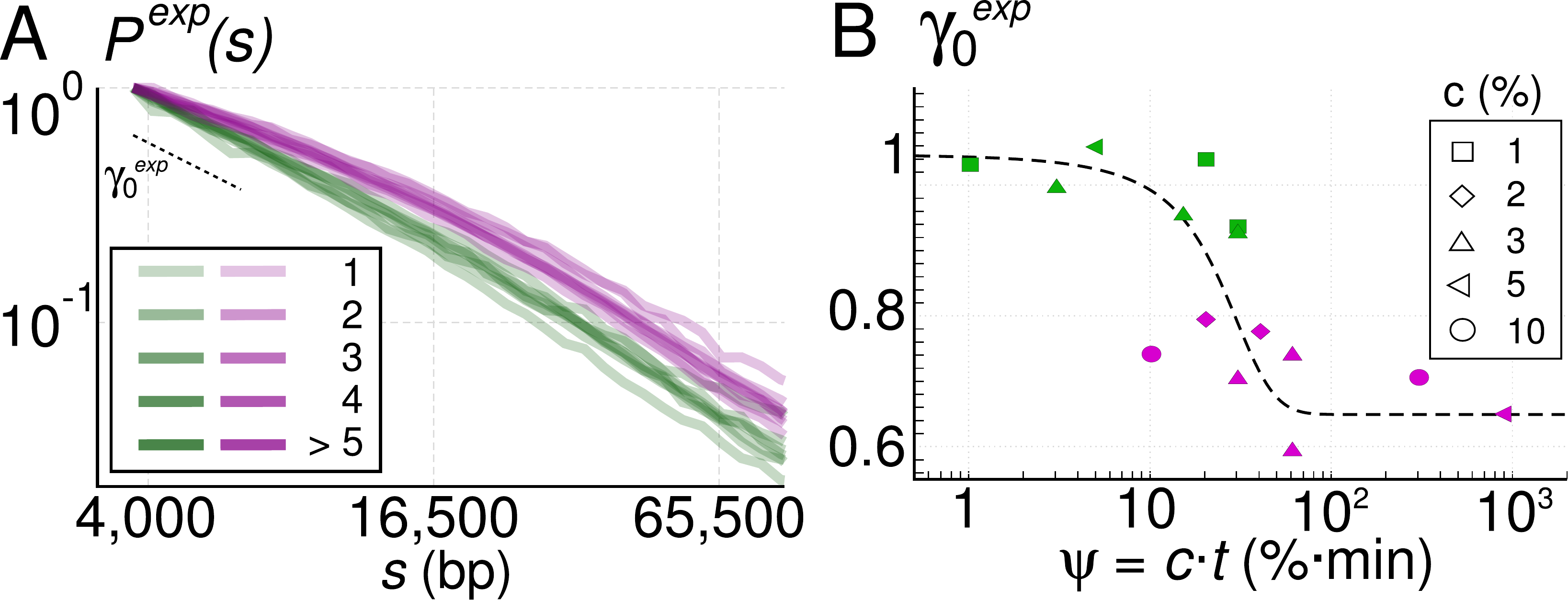}
    \caption{
      (A) Experimental contact probability curves $P^{exp}(s)$ for
      various crosslinker concentrations $c$,
      %for all experiments performed 
      displayed as a superposition of semitransparent plots (see Fig.~\ref{fig:first}A).
      (B) Evolution of the experimental slope $\gamma_0^{exp}$ as a
      function of the re-scaled time variable $\psi=c\cdot t$ (see
      Fig.~\ref{fig:first}B). The color discriminates the experiments
      belonging to the two modalities for $\gamma_0^{exp}$, the dashed line is a guide for the eyes.
    }
    \label{fig:exp}
\end{figure}

Experimental approaches in chromosome biology have been recently renewed by
%Regarding experiments, recent results obtained by 
 chromosome conformation capture (3C)  that uses a succession of
 crosslinking, restriction,
religation and sequencing steps to measure contact frequencies along a
DNA molecule {\it in vivo}.
This technique centrally exploits the unique opportunity offered by
the DNA heteropolymer to have a single sequence identifier at each
loci (for long enough identifiers) and so to derive  a contact
probability curve $P(s)$ from crosslink counts.
In the seminal paper introducing the genome-wide 3C technique, Hi-C, 
Lieberman-Aiden {\it et al.} ~\citep{lieberman2009comprehensive} 
%derived the contact probability curve $P(s)$ from crosslink counts, 
%as a function of the genomic 
%distance $s$. In the range between 1 and 10 millions base
%pairs (bp), these authors
 fitted the resulting curve with a scaling relation
$P(s) \propto s^{-\gamma}$, in the range between 1 and 10 millions base pairs (bp),
with a value of $\gamma$ close to $1$ compatible with a fractal-globule state.
 %proposed by Grosberg {\it et al}.
% Tension and loop extrusion
However, an  exponent of $0.75$ has also been reported at shorter scale, and other out-of-equilibrium mechanisms were
invoked to explain this alternative exponent: the tension
globule~\citep{sanborn2015chromatin} or the extrusion of loops by
molecular motors such as condensins ~\citep{sanborn2015chromatin, fudenberg2016formation}.
%terakawa2016condensin}.
While these mechanisms can have a role in chromosome folding, 
the models do not take explicitly into account the potential
distortion that the DNA polymer can undergo during the initial step of
the experiment, consisting in chemically crosslinking DNA with
formaldehyde.
This  crosslinking step prompted us to exploit this experimental
technique to check the collapse scenario described in our simulations.  

In order to start from configurations that are the closest possible
to a simple homopolymer, we used synchronized yeast cells
that are neither replicating nor dividing.
We performed Hi-C
(methods in SM
% MANUALREF
\S I.D and \citep{cournac2012normalization, lazar2017cohesins})
at different formaldehyde concentrations
$c$ and exposure times $t$
in order to observe
the evolution of polymer conformations during the crosslink-induced collapse.
Not knowing the reaction order, we cannot establish an exact  mapping
between $k_{on}$ and $c$, so
%need further investigations, as such we use an
we used a simple ansatz, $\psi = c \cdot t$, for the re-scaled  time
variable. The experimental curves $P^{exp}(s)$ cluster
around two different mean-curves differing by their slope at short
distances $\gamma_0^{exp}$ (Fig.~\ref{fig:exp}A). Plotting this
exponent as a function of $\psi$, we observe a sharp
transition~(Fig.~\ref{fig:exp}B) as predicted by the simulations.
Two differences are nevertheless worth discussing.
Before the transition, the short-distance exponent of yeast
chromosomes is not equal to $1.5$ as in simulations
(Fig.~\ref{fig:first}B), but to $1$ ($0.05 \;s.d.$). This value might either
correspond to an effect of volume interactions during the early phases
of pearling collapse or to an {\it in-vivo} special organization of
the DNA in chromosomes, potentially induced by the regular wrapping of DNA around the nucleosomal protein spools.
For distances above 10~kb these
constraints weaken and the chain follows a more typical random walk
with an exponent closer to $1.5$.
After transition, $\gamma_0^{exp}$ equals to $0.7$ ($0.06\;s.d.$), corresponding to the value observed for $N_{max}=1$ in simulations.
This value is likely explained by strong steric constraints preventing a crosslinked locus to contact other loci.
The precise estimation of $\xi$  was impaired by the higher biological, experimental and 
statistical noise on $P^{exp}(s)$ at increasing distance $s$, so that we could not measure experimentally the dependency 
of $\xi$ on the crosslinker concentration.
Nevertheless, the experiment clearly demonstrate that a polymer experiencing a crosslink-induced collapse
undergoes a sudden transition.
It also confirms that inside pearls, at length scales lower than
$\xi$, the polymer conformation in the absorbing asymptotic state is very
compact, with an  exponent $\gamma_0$ lower than $1$, whereas  the
polymer topology remains unchanged at longer length scales.

\vskip 2mm
We thank Madan Rao, John Marko, Jean-Marc Victor,
Benjamin Audit, Marco Cosentino-Lagomarsino, Maxim Dolgushev and Daniel Jost for the extremely
useful discussions and suggestions, and V\'eronique Legrand and the DSI
of Institut Pasteur for the computational power and assistance.

\let\oldaddcontentsline\addcontentsline% Store \addcontentsline
\renewcommand{\addcontentsline}[3]{}% Make \addcontentsline a no-op
\bibliography{biblio}
\let\addcontentsline\oldaddcontentsline% Restore \addcontentsline

% \usepackage[utf8]{luainputenc}
% \usepackage{latexsym} 
% \usepackage{amsmath}
% \usepackage{amssymb}
% \usepackage{booktabs}
% \usepackage{mathtools}
% \usepackage{setspace}
% \usepackage{color}
% \usepackage{verbatim}
% \usepackage{graphicx}
% \usepackage{placeins}
% \usepackage{wrapfig}
% \usepackage{fancyhdr}
% \usepackage{natbib}
% \usepackage[hidelinks]{hyperref}
% \usepackage{siunitx}
% \DeclareSIUnit\molar{\mole\per\cubic\deci\metre}
% \DeclareSIUnit\Molar{\textsc{m}}
% \usepackage[nokeyprefix]{refstyle}
% \usepackage{varioref}
% \usepackage{xr-hyper}
% \newcommand{\rev}[1]{\textcolor{red}{#1}}
% \newcommand*{\nolink}[1]{%
%   \begin{NoHyper}#1\end{NoHyper}%
% }

\clearpage

\renewcommand{\figurename}{Suppl. figure}
\renewcommand{\tablename}{Suppl. table}

\clearpage
~
\vspace{7cm}\\
~
{\center\Huge
Supplementary Materials\\
}
\clearpage
\tableofcontents

\setcounter{figure}{0}
\section{Methods}

\subsection{Simulation method}

\subsubsection{Source code}

The code for reproducing the simulations contained in this manuscript
is available on GitHub at
\url{https://github.com/scovit/crosslink} and is compatible with the
Linux operating system. Compilation requires a recent version of gcc,
GNU make, OpenGL,
OpenSSL, flex and bison and can be achieved by the command
\verb|make|. Hardware requirements include a recent x86-64 CPU with
supports for the AVX instruction set. After compiling, the code can be
run with the command \verb|./crosslink[.gl] configuration.info|.
Sample configuration files are provided in the \verb|samples|
folder and the optional extension \verb|.gl| activates the real-time
graphical visualization of the simulation.

\subsubsection{Algorithm description}
We simulate the crosslinking process under the
minimal assumptions that (1) chromosomes are ideal chains of monomers
and (2) the effect of crosslinking is an irreversible topological
change between distant monomers on the chain.
We model a chromosome as a $2048$ beads polymer fluctuating in three
dimensional space using a variant of the off-lattice bead-spring
Monte-Carlo (MC) algorithm described in refs.
\citealpltex{cacciuto2006self, scolari2015combined}.
Each bead of the polymer represents a group of $3$ nucleosomes, or
$500$ base pairs (as in ref. \citealpltex{hajjoul2013high}), adjacent
beads are linked by an infinite spherical potential of radius
$b$, thus forbidding any dynamical moves which would settle them
at larger distances while allowing everything else, each MC move
selects randomly a single bead and attempts a move in a random
direction uniformly, normally sampling a sphere of radius
$0.6 \cdot b$. Different mean-square displacements (MSD) for a bead   have been obtained by varying the size
of the random move from $0.066 \cdot b$ to $0.2 \cdot b$. 

We used the value of
$b = 24$nm, which is an estimation of the maximum distance that can
be covered by three fully stretched nucleosomes, and one MC sweep as
$6.7 \cdot 10^{-4}\,$secs, estimated by fitting the motility of the
tracking dynamics of single loci \citeltex{hajjoul2013high}, see next
section for details.
%
% The Rouse dynamics predicts a genomic length dependent relaxation time
% (see Appendix, before \nolink{\eqref*[xr=App-]{eq:relaxexact}}, for
% details):
% \begin{equation} \tau_R(s) =
% \frac{2}{\pi^2}\left(\frac{D_s}{D_t}\right)^2 \cdot s^2
% \end{equation}

% Where the expression of $D_s$ and $D_t$ is given in appendix and
% depends on these two scaling factors.
%
After a fixed thermalization time $t_0$, the crosslinking process is
introduced in the simulation as an irreversible and
configuration-dependent change in the chain topology.
After each accepted MC move, if a couple of beads are found at a
distance lower than $r_{int} = b / 64$ units a link is introduced with a
given probability $p$ that we consider as a proxy for the
crosslinker concentration $c$, and the couple is added to the list of
adjacent beads.
%, for rationales behind this choice, please refer to
%Appendix \nolink{\secref[xr=App-]{sec:interaction}}.
%
Crosslinks are totally irreversible, and a maximum number of
crosslinks $N_{max}$ for each bead is allowed, mimicking steric
effects. When it is not otherwise specified, $N_{max}$ is taken equal
to 4.
Such events are then counted as contacts and mean contacts maps are
built by summing all the contacts over a population of independently
simulated chains.

\subsubsection{Reproduction of the polymer equilibrium dynamics}
\label{sec:repeq}

The simulation integrates the stochastic Rouse dynamics as an
effective model using a purely entropic dynamical Monte-Carlo (MC) algorithm. This
section details the reproduction of Rouse dynamical features using the simulated dynamics.
Rouse Dynamics predicts two  power-law behaviors:
(1) the trajectory of a polymer unit segment (monomer) displays a mean
square displacement (MSD) scaling with time with a sub-diffusive
exponent $1/2$, and
(2) at any fixed time, the conformation of the polymer displays a mean
square distance scaling with the backbone distance with an exponent $1$. The
coefficients $C_t$ and $C_s$ are defined from these scaling behaviors
according to:
\begin{equation}
\label{eq:coeffdef}
\left\langle \left| \vec R(s, t_0) - \vec R(s, t + t_0) \right|^2
\right\rangle = C_t \cdot t^{1/2}
\quad \mathrm{and} \quad
\left\langle \left| \vec R(s_0, t) - \vec R(s + s_0, t) \right|^2
\right\rangle = C_s \cdot s,
\end{equation}
where $\vec R(s, t)$ is the location at time $t$ of the monomer at
position $s$ along the polymer chain,
%trajectory of a realization of the polymer in time, 
$\langle \cdot \rangle$ is the mean over a
population/ensemble of chains, $t_0$ 
and $t + t_0$ are two observation times, and $s_0$ and 
$s + s_0$ are two genomic coordinates; for details and derivations
see \citeltex{doi1988theory} and Appendix, \secref{seq:relaxexac}.
In the experimental situation we considered, measurements of {\it
  in-vivo} Yeast chromatin reported the validity of
both scaling behaviors~\citeltex{bystricky2004long, hajjoul2013high},
and provide a direct measurement of the two parameters $C_t$ and $C_s$
characterizing the equilibrium dynamics of the system.

\begin{figure}[t]
    \centering
    \includegraphics[width=0.7\linewidth]{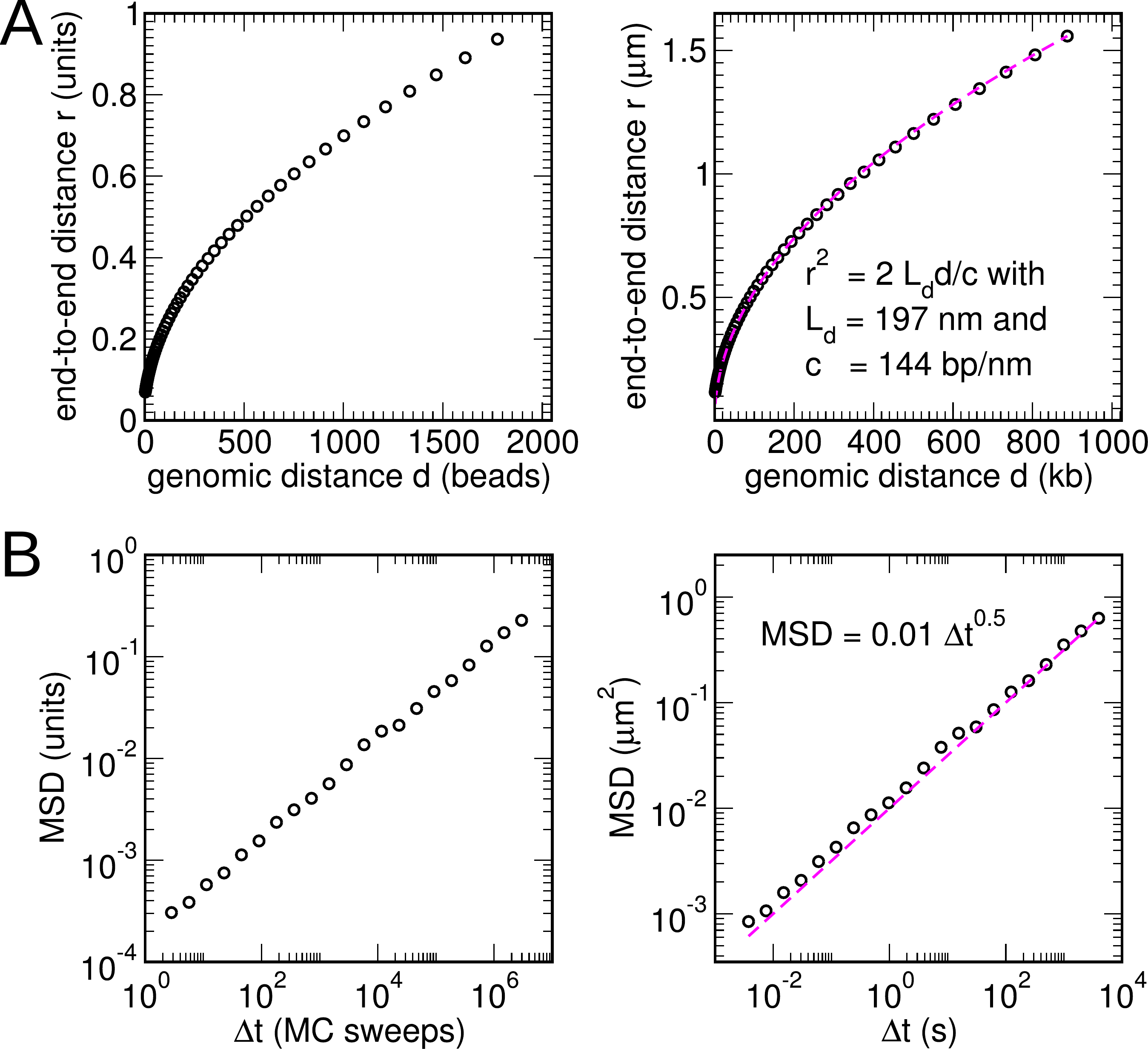}
    \caption{Rouse  dynamics and reproduction by the simulated dynamics of published
      experimental coefficients: (A)
% The unit of measure for space (${\rm \mu
%        m}$) has been 
%      defined from the 
      measurement of the end-to-end distance as a
      function of the genomic distance at equilibrium; the left panel
      displays the numerical data from simulations, the right panel displays
      the data rescaled to fit with the observed coefficient
      $C_s = 2 \cdot 197 / 144\ {\rm nm^2 / bp}$, adapted from
%      \citeltex{bystricky2004long}.
      (B) The unit of measure for time (s) has been defined by fitting
      the coefficient of power-law describing the MSD as a
      function of time with the observed motility coefficient
      $C_t = 0.01 {\rm \mu m^2 / s^{1.2}}$, adapted from
      % \citeltex{hajjoul2013high}.}
      }
    \label{fig:fittone}
\end{figure}

It is possible to assign units of time ($sec$) and
space ($nm$) to simulated quantities by relating the numerical
coefficient $C_t$ and $C_s$ in the simulations and their values
in the experimental results. The proof of concept is presented in Suppl. \figref{fig:fittone}, demonstrating that
% which shows the
%reproduction of 
it is possible to fit the experimentally measured  behaviors and associated coefficients with those observed in the simulation.
%numerical coefficients by measuring
%the MSD in function of time and the mean distance in function of
%backbone distance in simulations.
%
While this approach could allow us to fix the values of the
biophysical units using the most up-to-date literature,
we acknowledge that this approach presents the flaw that those two
parameters depends on the experimental conditions and scale, see
\citeltex{grosberg2016scale} and Appendix,
\secref{seq:relaxexac}. Depending on the
experimental conditions, the scale value can correspond to totally
different microscopical
quantities, namely it can depend on the size of the fluorescent
locus during optical measurements~\citeltex{javer2013short}, the length of
the DNA-linkers between nucleosomes in {\it in-vitro}
experiments~\citeltex{bednar1998nucleosomes, kornberg1977structure}, a
characteristic length dependent on the local density of 
monomers if the excluded-volume is predominant~\citeltex{wyart2000viscosity, brochard1979conformations}, a characteristic
time dependent on active noise fluctuations in dynamical
measurements~\citeltex{weber2012nonthermal, osmanovic2017dynamics}, or
other constraints and microscopical properties of the chromatin, for
instance \citeltex{weber2010bacterial, amitai2013polymer}. Additionally,
some of these elements~\citeltex{weber2010bacterial, amitai2013polymer,
  weber2012nonthermal, osmanovic2017dynamics}, as well as strong
steric~\citeltex{marenduzzo2006entropy}
and hydrodynamic~\citeltex{doi1988theory} effects, can alter the
scaling behaviors of \eqref{eq:coeffdef}. The two
experiments  reproduced in Suppl. \figref{fig:fittone}
(\citeltex{bystricky2004long, hajjoul2013high}) display a robust
scaling, but are not made in identical controlled environments. As
such,
% It does not
%  The Appendix
% \secref{seq:relaxexac} enters in more details on this issue, including
% providing an alternative strategy
% % Using the time rouse
% for the solution to this problem eventually exploitable in future
% works.
we decided to fix the simulation spatial unit through a
definition of the maximum extension of linkers between 
three nucleosomes $b = 24$nm and, after, to fit the temporal
 scaling behavior of MSD as in
Suppl. \figref{fig:fittone}B for the definition of the unit of
time.

To obtain Suppl. \figref{fig:fittone}, the end-to-end distance as a
function of the genomic distance, we measured the equilibrium
conformation along time of 64 parallel polymer simulations, then
calculated the mean distance in function of genomic distance and
log-binned the data though a geometric progression of ratio of 2.
To obtain the MSD as a function of time, we collected for the same 64 parallel
equilibrium simulations the trajectories of the bead placed in the
center of the polymer chain; then we subtracted from each trajectory
the position at time zero and we calculated the square of each
three-dimensional vector and log-binned the resulting curve through a
geometric progression or ratio of 2.

\subsubsection{Irreversible crosslinking process}
\label{sec:interaction}

% Cutted part of main text
%In this view, the pearling transition corresponds to a transition
%between a process of reaction or diffusion-limited aggregation
%depending on the scale set by the genomic distance in the $p(s)$.
%

The simulation implements Rouse dynamics with the addition of
crosslinks, namely irreversible and
configuration-dependent changes in the chain topology.
In detail, after each accepted MC move, in case a couple of beads are found
at a distance lower than $r_{int} = b / 64$ units, a crosslink is
introduced between them with a given probability $p$ that we consider
as a proxy for the crosslinker concentration $c$ and the couple is
added to the list of adjacent beads (in the 3D space).
This section reconsiders this proxy connecting the parameter
$p$ to physico-chemical quantities.
%  and relations that we do not consider in
% the current work, but that opens the possibility to explore the
% current results under the optic of a more general and scale-dependent
% reaction to diffusion limited aggregation (DLA) phase transition.

The problem of  irreversible reactions can be
restated as a first passage-time problem: in the most basic examples
(e.g. the decay radioactive atomic nuclei), the on-rate of
transition from state A to state B is a constant in time, which is
called $K_{on}$. As such,
considering a finite (continuous) amount of time $\Delta t$, the probability
$p$ of passing from A to B in that amount of time can be
calculated from the rate of doing the transition
integrated over the elapsed time
\begin{equation}
  p = 1 - e^{- K_{on} \Delta t}.
\label{eq:pk}
\end{equation}
%with $p\approx K_{on} \Delta t$ when $K_{on} \Delta t\ll 1$.
%
% Numerical simulations can
% be expressed by an algorithm with definite timesteps $\Delta t$ and
% repeated tests with probability $p$, increasing
% the total ellapsed time, or alternatively by a timestep-free
% algorithm such as Gillespie\citeltex{erban2007practical} which
% stochastically generate the time of reaction.

%
The general problem of  polymer crosslinking reaction  is more complex because at least two concurrent
processes are contributing to the probability of making a crosslink:
(1) the diffusion of the
polymer is creating three-dimensional contacts, that is a necessary condition
for the reaction to happen and (2) the reaction of the crosslinking
agent with proteins and then with DNA fixes the contacts defined that
way. The first process is simulated by the Rouse dynamics
described in the previous section, i.e. contacts get continually
activated and dissolved depending on the polymer parameters and the
genomic distance. The second process, described in chemical details in
Hoffman et al.~\citeltex{hoffman2015formaldehyde}, is, in our simulations,
considered as an
irreversible transition parameterized by a single parameter $p$
depending on crosslinker concentration and phenomenologically modeling
the detailed description.
%
%  of the diffusion of the
% proteins in the cytoplasm and/or the crosslinking agent as well as the
% correct modelisation of the chemical steps which allow the crosslinking
% to happen and the relative rates, as we do, corresponds
%
% In fact, making the
% implicit hypotesis that the polymer dynamics is much slower ($k_{on}$,
% and thus $p$ is small in comparison to invers of the smallest
% relaxation times), and, as such, that the interesting events happens
% only at the mesoscales described by our Rouse dynamics.
%
%

\begin{figure}[t]
    \centering
    \includegraphics[width=0.9\linewidth]{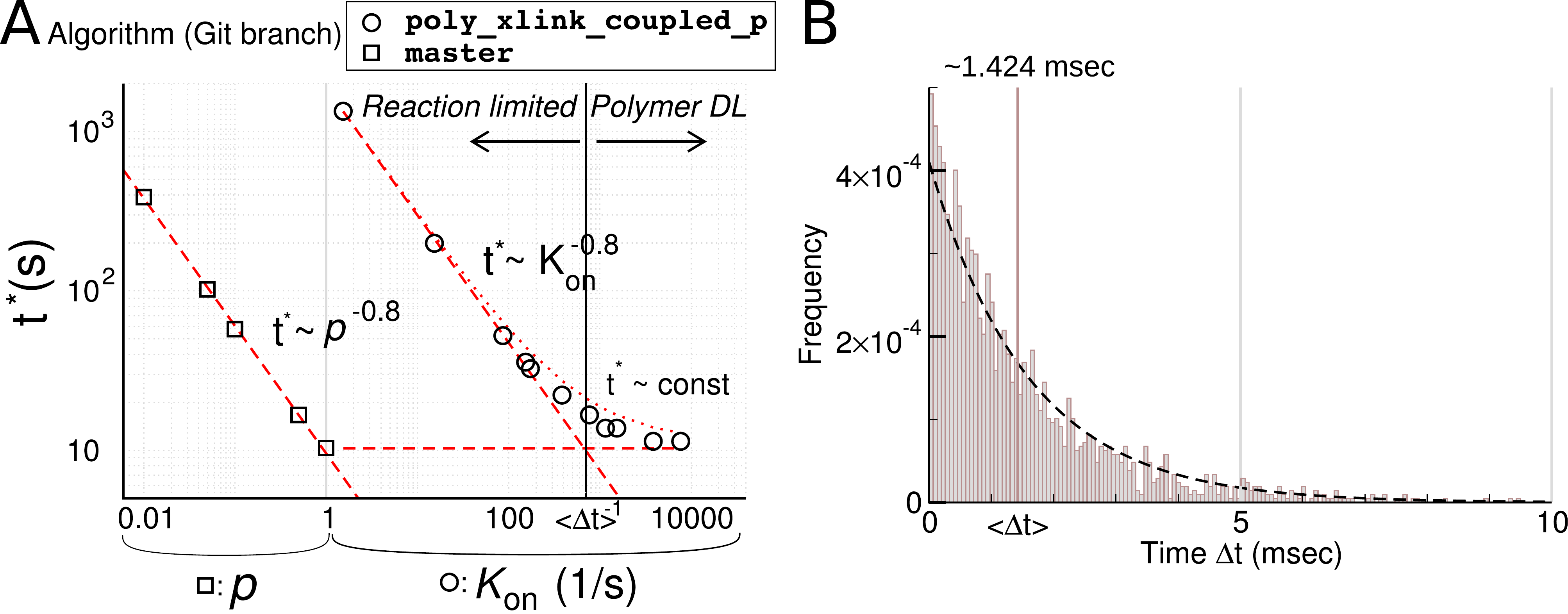}
    \caption{Transition between reaction-limited to diffusion-limited 
      crosslinking dynamics: (A) Results from simulations using the
      \texttt{poly\_xlink\_coupled\_p} algorithm (circles) show a scaling relationship with exponent $0.8$ between the transition time $t^*$
      and the association constant  $K_{on}$ in the reaction-limited regime 
%      critical ?
       (dashed red lines); they also evidence the transition for
      $\langle \Delta t \rangle \gg 1/K_{on}$ to a polymer
      diffusion-limited regime (DL). Results from simulations using the
      \texttt{master} algorithm
      (boxes) display a similar scaling in the reaction-limited
      regime upon  rescaling $K_{on}$ by the transition value
      $\langle \Delta t \rangle^{-1}$. (B)  The 
      distribution of the duration $\Delta t$ of contacts 
      in our algorithms depends on the Rouse chain parameters $C_t$ and
      $C_s$, and it might depend on additional
      biophysical constraints in experiments. The histogram presented
      here has been obtained
      with the same parameters as those used
      in the main text.
    }
    \label{fig:difflim}
\end{figure}

A study of the simple function in \eqref{eq:pk} reveals that
if $\Delta t$ is small ($\Delta t \ll 1/K_{on}$), then the dependency
between the probability $p$, the rate $K_{on}$ and time $\Delta
t$ can be approximated by this linear relation:
$p \simeq K_{on} \Delta t$.
Since the distribution of $\Delta t$ in our simulations is
short-tailed (see Suppl. \figref{fig:difflim}B), we can take
its mean as a value for $\Delta t$ in \eqref{eq:pk},
connecting the parameter $K_{on}$ with the dimension of a rate
(inverse of a time) to a crosslink probability $p$. The results of
this approach are
valid as long as the simulation is in this reaction-limited
regime ($K_{on} \ll 1/\Delta t$ such that $p \ll 1$) at the
lowest spatial scales.
%, corresponding to the cutoff of three nucleosomes for bead.

%
Regarding the connection between $K_{on}$ and the crosslinker concentration $c$, the
emergence of a single experimental collapse curve for the
exponent $\gamma_0$ when it is considered as a function of the rescaled time variable 
$\psi = c \cdot t$ (see
% MANUALREF
figure 4B, main text)
suggests that the transition rate $K_{on}$ depends linearly on the
concentration, although from this single experiment, we cannot
exclude reaction orders smaller than 3. From the biophysical point of
view, the dimensionless values of $p$ from simulations could be related
quantitatively to experiments once a precise quantitative measurement
of the reaction order and a quantitative measurement of the typical contact duration $\Delta t$ would have been performed.

An alternative, but more complex, approach (code available in the
\verb|poly_xlink_coupled_p| branch in GitHub) to relate the crosslinking dynamics to the polymer dynamics,
consists in having the simulation keeping track of the time a
contact has been created, and, at the time when the contact is
released, calculate the probability of having made a crosslink
during that time as the discrete version of \eqref{eq:pk} above:
$p = 1 - (1 - k_{on})^{\Delta t}$ where $\Delta t$ is the discrete
elapsed time expressed in terms of number of MC steps and $k_{on}$ is a dimensionless rate;
%expressed in units of MC steps inverted; 
then, according to a
uniform sampling, decide if a crosslink has been made or not.
A study of the results reveals the presence of a transition
between a reaction-limited regime and a polymer-diffusion-limited
regime:
if $\Delta t$ is small ($\Delta t \ll 1/k_{on}$), then the system is
in a reaction-limited regime, and the dependency between the probability
$p$, the rate $k_{on}$ and time $\Delta t$ can be approximated by
this linear relation:
$p \simeq k_{on} \Delta t$. For larger $\Delta t$,
 the probability is instead equal to $p \simeq 1$, and the
crosslink reaction becomes only polymer-diffusion-limited. We plotted
the transition curve for $\gamma_0$ with this algorithm and show the
presence of the two distinct behaviors for the dependency of the transition time on $k_{on}$:
a power law with exponent $0.8$ in the reaction-limited regime (see main text) and no dependency on $k_{on}$ in the
diffusion-limited regime (see Supplementary
\figref{fig:difflim}A, circles). The results, apart from a rescaling,
are similar to the results of the simpler algorithm (Supplementary
\figref{fig:difflim}A, squares).

%The emergence of an anomalous exponent $\delta = 0.4$ can be seen
%as a signature of universality and the consequence of another constant
%of dimension of time, that we identify as the mean microscopical
%duration of physical contacts at equilibrium, $\tau_{DLA}$, and is
%fixed in our simulations. Using this property, we can express the
%probability of making a crosslink in term of a reaction rate
%$r = p / \tau_{DLA}$, in this extension, $p$ can be any positive value
Finally, we speculate that taking into account the transition from
diffusion-limited to reaction-limited regimes, the expressions for pearling
size $\xi$ and time $t^*$ will be (denoting $\xi_{cr}$ and
  $t^*_{cr}$ the pearling size and  time for $p=1$):
\begin{equation}
  \begin{split}
  & \frac{t^* - t^*_{cr}}{t^*_{cr}} = p^{-2\delta},
  \  \  \    \mathrm{and}          \\
  & \frac{\xi - \xi_{cr}}{{t^*_{cr}}^{1/2}} = D_R^{1/2} p^{-\delta},
  \  \  \    \mathrm{with}         \\
  & \xi_{cr}^2 = D_R \cdot t^*_{cr} \ \mathrm{and}\ \delta = 0.4;
  \end{split}
  \label{eq:xit2}
\end{equation}
which model the crossover.
Equation \eqref*{eq:xit2} resumes equations
% MANUALREF
5, 6, 7 and 8
from the main text. They
display explicitly the dependency on the crosslink probability $p$,
introduce a soft crossover toward the critical value $p = 1$ and the bounds
$\xi_{cr}$ and $t^*_{cr}$ to be determined experimentally. 
%At the moment, our experiments are not designed to study thetransition in this specific detail.
%, also, a more careful theoretical
%analysis
% taking into the renormalization flux in the space and time
% values involved in the observation of this transition
%will be necessary to verify this speculation.

\subsection{Simulation data analysis}
\label{sec:simana}

\subsubsection{Generation of contact matrices}

\begin{figure}[h]
    \centering
    \includegraphics[width=\linewidth]{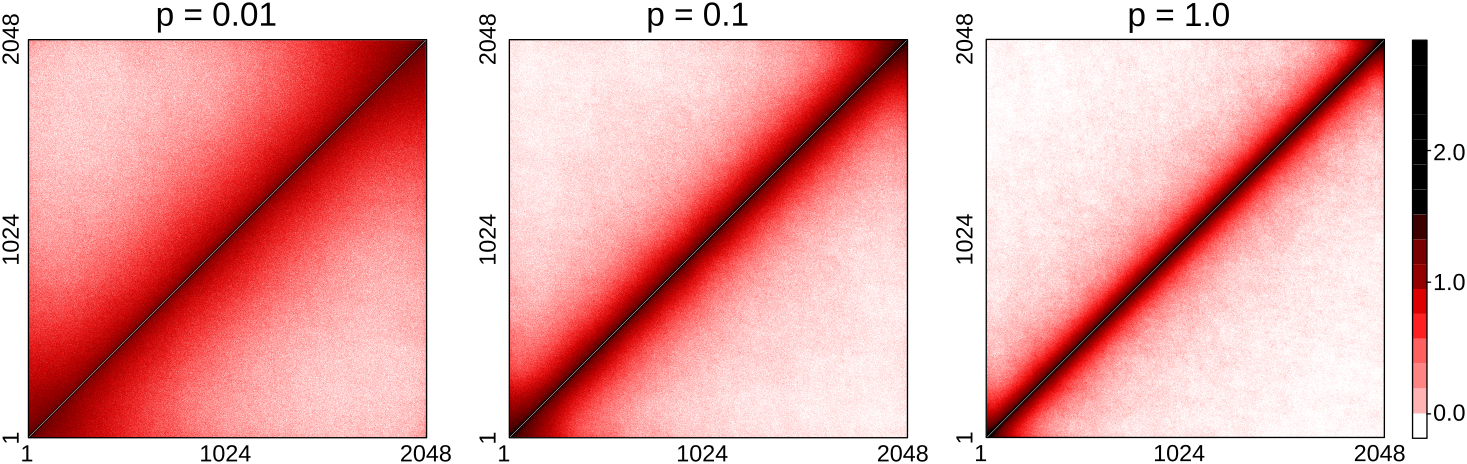}
    \caption{Contact matrices for ideal polymer models (simulation) at very long
      time (after the transition) at various crosslinking probabilities}
    \label{fig:contmat}
\end{figure}

For each set of parameters, simulations has been launched in batches of
2000 identical runs on the \verb|TARS| cluster of ``Pasteur
Institut'', powered by 3500 identical CPUs. Time and information
about the couple of involved beads have been recorded for each
crosslinking event for each parallel simulation in distinct
data-files.
Sparse matrices have been generated, merging all batches, summing
the number of crosslinking events happening for each couple of beads up to
a specified time. Time has been sampled according to a geometric
progression of initial value and common ratio equal to $1.1$,
discarding all matrices with less than $512$ contacts.
Contact matrices for ideal polymer models at very long
time (after the transition) at various crosslinking probabilities
look like the one reported in Suppl. \figref{fig:contmat}.

\subsubsection{Calculation of $P(s)$ in the simulation}

In the simulation, sparse matrices have been coarse-grained to bead-level dense matrices. Binning of the contact probability
$P(s)$ at bead level has been made by calculating the mean
value of secondary diagonals  for $s$  varying in the range from 1 to 2048 beads.  
In  order to reduce the noise level at large linear distances $s$ (along the polymer chain) and
avoid the effect of chain discretization at small  distances,
the curve $P(s)$ has been binned according to the following scheme: values
for $s$ less then $5,000$bp were discarded; then binning has been performed at single bead level
(bins of $500$bp) for distances $s$ less than $10,000$bp, then we used bins
of $1,000$bp for $s$ between $10,000$ and $20,000$bp, bins of $2,000$bp for
$s$ between $20,000$ and $40,000$bp, and finally bin sizes defined according
to a geometric progression (log-binned) of initial value and common
ration equal to $1.1$ for  distances $s$ greater than $40,000$bp.
The contact distribution  is obtained as
the weighted histogram computed from the sum of read pairs for
each bin, weighted with the expected number of
pairs under the uniform null hypothesis, which takes into
consideration the bin size. To compare different
conditions, distributions were normalized by the value of their first
bin, yielding contact probability curves $P(s)$.

\subsubsection{Determination of the collapse transition}

The local slope $\gamma(s)$ along the chain, at a linear distance $s$,  has been
calculated by performing a linear regression on a sliding-window on
the log-log plot of the $P(s)$, using a window of size equal to 7
data-points. The collapse transition has been determined  by 
collecting, for each simulation condition, the slope
 $\gamma$ of the linear regression performed on the first 7
data-points falling in the interval  of $s$ values between $5,000$ and $8,500$ bp.

\subsubsection{Determination of the transition values $t^*$ and $\xi$}

The pearling size $\xi$ has been measured on the local slope $\gamma(s)$
as a function of the linear distance $s$ by measuring the first value  $s$ for which $\gamma(s) > 0.9$.
The transition time $t^*$  (pearling time) has been measured  as the first value of
time $t$ for which $\gamma_0 > 0.9$.

\subsubsection{Determination of the Rouse coefficient from MSD scaling}
\label{sec:rousemsd}

The Rouse coefficient, $D_R$, has been calculated in the simulation from the measured prefactor, $C_t$,  in the 
scaling of the MSD as a function of time  and the effective
persistence length, $C_s$,  through the formula (see Suppl. Methods
\secref{sec:repeq}) 
\begin{equation}
  D_R = \frac{\pi^3}{4} \left( \frac{C_t}{C_s} \right)^2,
\end{equation}
see Appendix \secref{sec:relaxexact} for the derivation.

\subsection{Generation of Hi-C data-sets}

\subsubsection{Crosslink conditions on G1 elutriated cells}

{\it S. cerevisiae} [BY4741] cells were inoculated and for
\SI{8}{\hour} in \SI{10}{\mL} YPD, then \SI{500}{\uL} of the
pre-culture were inoculated and grown 
overnight in \SI{500}{\mL} YPD. \SI{500}{\mL} overnight culture was
centrifuged and pelleted, then cells were resuspended in 
\SI{500}{\mL} of fresh YPD for \SI{3}{\hour} at
\SI{30}{\degreeCelsius}. G1 daughter cells were recovered from this 
exponentially growing population through an elutriation procedure
\citeltex{marbouty2014purification}. Before fixation, G1 cells were
refreshed in \SI{150}{\mL} of fresh YPD at \SI{30}{\degreeCelsius} for
\SI{30}{\minute} ($2.5 \times 109$ G1 cells/fraction). Cells were
crosslink using 
formaldehyde (Sigma) under different conditions of concentration and
time (see Suppl. Table \ref{tab:xlc}). The crosslink reaction was
quenched with \SI{25}{\mL} glycine \SI{2.5}{\Molar} for
\SI{20}{\minute} at \SI{4}{\degreeCelsius}. Crosslinked cells were 
recovered through centrifugation, washed with YPD, pelleted and stored
at \SI{-80}{\degreeCelsius} into \SI{2}{\mL} centrifugal tube.

\subsubsection{Generation of Hi-C libraries}

Hi-C libraries were generated as described in
\citepltex{dekker2002capturing, cournac2016generation} with introduction of
a biotin-ligation
step in the protocol \citepltex{lieberman2009comprehensive}. To generate
the libraries, a pellet of G1 cells, previously crosslinked, was
thawed on ice. Then, the cells were incubated for
\SI{30}{\minute} in \SI{10}{\mL} of sorbitol \SI{1}{\Molar} with DTT
\SI{5}{m\Molar} and Zymolyase 100T (CFinal = \SI{1}{\mg/\mL}) to
digest the cell wall. Spheroplasts were then washed first with
\SI{5}{\mL} of sorbitol \SI{1}{\Molar}, then with \SI{5}{\mL} of 1X
restriction buffer (depending on the restriction enzyme
used).
Spheroplasts were washed with \SI{5}{\mL} sorbitol
\SI{1}{\Molar}, then with \SI{5}{\mL} 1X restriction buffer (NEB), and
suspended in \SI{3.5}{\mL} 1X restriction buffer. Cells were split
into aliquots (V = \SI{500}{\uL}) and incubated in SDS (3\%) for
\SI{20}{\minute} at \SI{65}{\degreeCelsius}. Crosslinked DNA was
digested at \SI{37}{\degreeCelsius} overnight with 150 units of DpnII
restriction enzyme (NEB). The digestion mix was subsequently
centrifuged for \SI{20}{\minute} at 18,000 g and the supernatant
discarded. Pellets were suspended in cold water.
DNA ends were
repaired in the presence of 14-dCTP biotin (Invitrogen), and
crosslinked complexes incubated for \SI{4}{\hour} at
\SI{16}{\degreeCelsius} in presence of 250 U of T4 DNA ligase (Thermo
Scientific, \SI{12.5}{\mL} final volume). DNA purification was
achieved through an overnight incubation at \SI{65}{\degreeCelsius}
with \SI{250}{\ug/\mL} proteinase K in \SI{6.2}{m\Molar} EDTA followed
by a precipitation step and RNAse treatment.
The resulting Hi-C DNA libraries were 500 bp fragmented, using
CovarisS220 apparatus. Fragments between 400 and 800 bp were purified
and the biotin-labeled fragments were selectively captured by
Dynabeads Myone Streptavidin C1 (Invitrogen). Purified fragments were
amplified by PE-PCR primers and paired-end sequenced on the NextSeq500
Illumina platforms ($2 \times 75$ bp).

\subsubsection{Raw data processing}

\label{sec:hicgen}

Suppl. \tabref{tab:xlc} contains the information about experimental crosslinking
conditions.

\begin{table}[h]
  \begin{tabular}{l|c|r|r}
{\bf strains} & {\bf crosslink concentration} &
{\bf crosslink time} & {\bf \# reads}\\
\hline
BY4741 & 1 \% (V = 4.2 mL) & 1 min & 9190059 \\
BY4741 & 1 \% (V = 4.2 mL) & 20 min & 1608139 \\
BY4741 & 1 \% (V = 4.2 mL) & 30 min & 13693706 \\
BY4741 & 2 \% (V = 8.4 mL) & 10 min & 3056592 \\
BY4741 & 2 \% (V = 8.4 mL) & 20 min & 5655336 \\
BY4741 & 3 \% (V = 12.6 mL) & 1 min & 1977469 \\
BY4741 & 3 \% (V = 12.6 mL) & 5 min & 1616825 \\
% MeiosynIV_Crosslink_3%_10min
BY4741 & 3 \% (V = 12.6 mL) & 10 min & 6091165 \\
% BY_Crosslink_3%_10min
BY4741 & 3 \% (V = 12.6 mL) & 10 min & 1630060 \\
% BY_Crosslink_3%_20min
BY4741 & 3 \% (V = 12.6 mL) & 20 min & 1389810 \\
% BY_Crosslink_3%_20min_bis
BY4741 & 3 \% (V = 12.6 mL) & 20 min & 2079929 \\
BY4741 & 5 \% (V = 21 mL) & 1 min & 2223222 \\
BY4741 & 5 \% (V = 21 mL) & 180 min & 1143955 \\
BY4741 & 10 \% (V = 42 mL) & 1 min & 18367276 \\
BY4741 & 10 \% (V = 42 mL) & 30 min & 17466440 \\
  \end{tabular}
\caption{
 Genomic DNA in living cells was crosslinked using formaldehyde (Sigma) under different
  conditions of  crosslinker concentration and experiment duration
}
\label{tab:xlc}
\end{table}

Raw Hi-C data were processed as follows. PCR  duplicates were removed
using the 6 Ns present on each of the custom-made adapter and the 2
trimmed Ns. Paired-end reads were mapped independently using Bowtie
2.1.0 (mode: \verb|--very-sensitive --rdg 500,3 --rfg 500,3|) against
the {\it S. cerevisiae} reference genome (S288C). An iterative
alignment, with an increasing truncation length of 20 bp, was used to
maximize the yield of valid Hi-C reads (mapping quality $> 30$). Only
uniquely mapped reads were retained. On the basis of their DpnII
restriction fragment assignment and orientation, reads were classified
as either valid Hi-C products or unwanted events to be filtered out
(i.e., loops, non-digested fragments, etc.); for details see
\citeltex{cournac2012normalization, cournac2016generation}. The amount of
reads in contact maps is reported in table \tabref{tab:xlc}.

\subsection{Experimental data analysis}
\label{sec:hicana}

\subsubsection{Calculation of $P^{exp}(s)$ from Hi-C data}

\begin{figure}[h]
    \centering
    \includegraphics[width=0.7\linewidth]{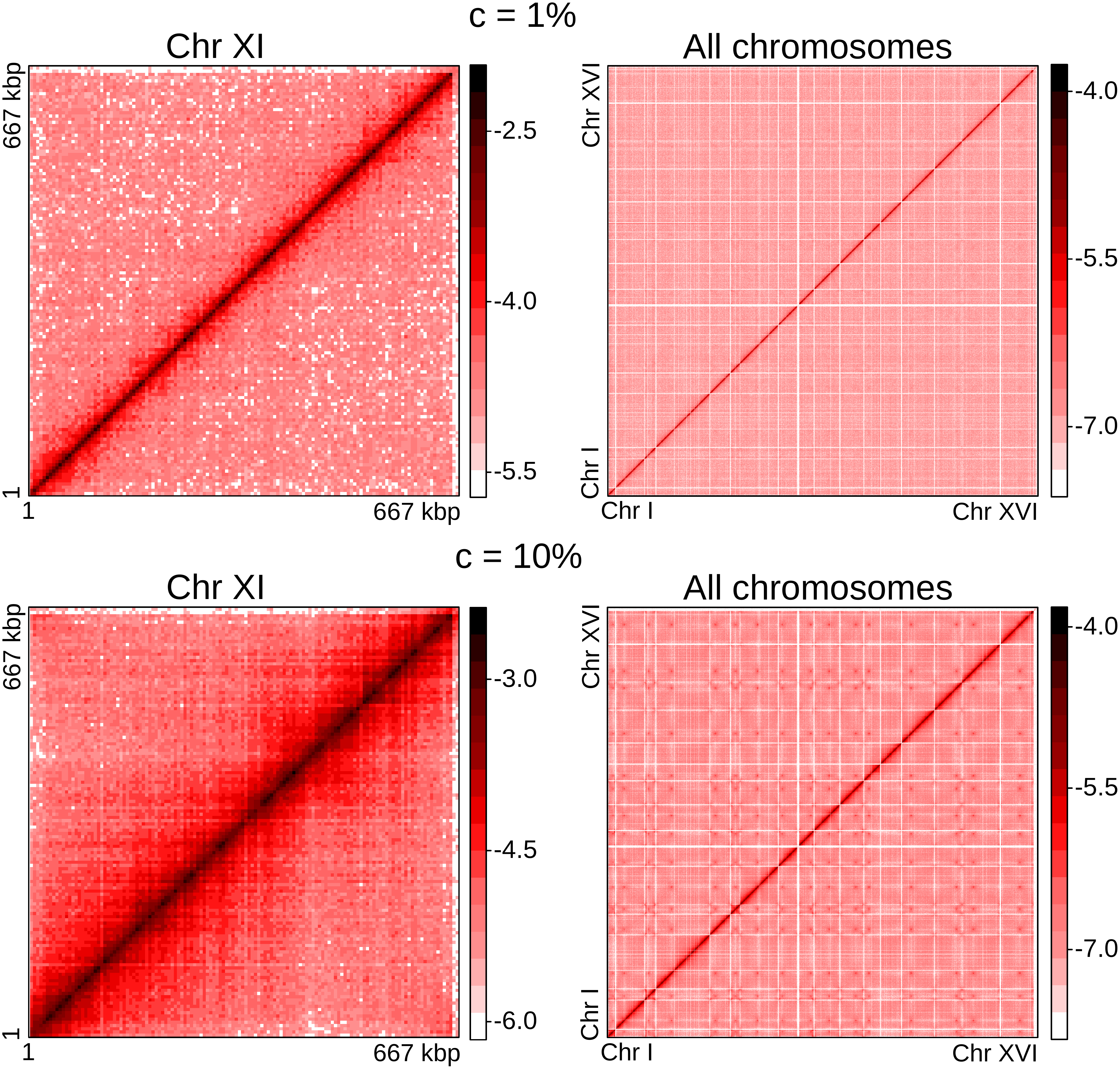}
    \caption{Sample experimental contact matrices before and after the
      transition}
    \label{fig:expmat}
\end{figure}

Pairs of intra-chromosomal reads mapping positions along
the genome were partitioned according to chromosomal arms. Reads oriented towards
different directions or separated by less than 3~kb were
discarded. For each chromosomal arm, except for the right arm of the
chromosome XII which comprises the rDNA, read pairs were log-binned
according to the genomic distance $s$ separating them (in kb),
$bin = \mathrm{floor}[ \log_{1.1}(s) ]$. The contact probability
distribution $P^{exp}(s)$ is
the  histogram computed from the sum of read pairs for
each bin, locally normalized by the expected number of
pairs in this bin under the uniform null hypothesis. To compare different
conditions, distributions were globally normalized by the value of their first
bin.
Sample experimental contact matrices before and after the transition
are reported in Suppl. \figref{fig:expmat}.

\subsubsection{Determination of the collapse transition}

The local slope $\gamma^{exp}(s)$ as a function of the genomic distance $s$ has been
calculated by performing a linear regression on a sliding-window on
the log-log transformation of $P^{exp}(s)$, on a window of the size of 7
data-points. The collapse transition has been built by 
collecting, for each crosslink concentration and time, the slope
coefficient of the linear regression performed on the 7 data-points
falling in the $s$ interval between 3,800~bp and 6,150~bp.

\clearpage
\section{Appendix}

\subsection{Measurement of pearls with Louvain algorithm}

\begin{figure}[h]
    \centering
    \includegraphics[width=0.7\linewidth]{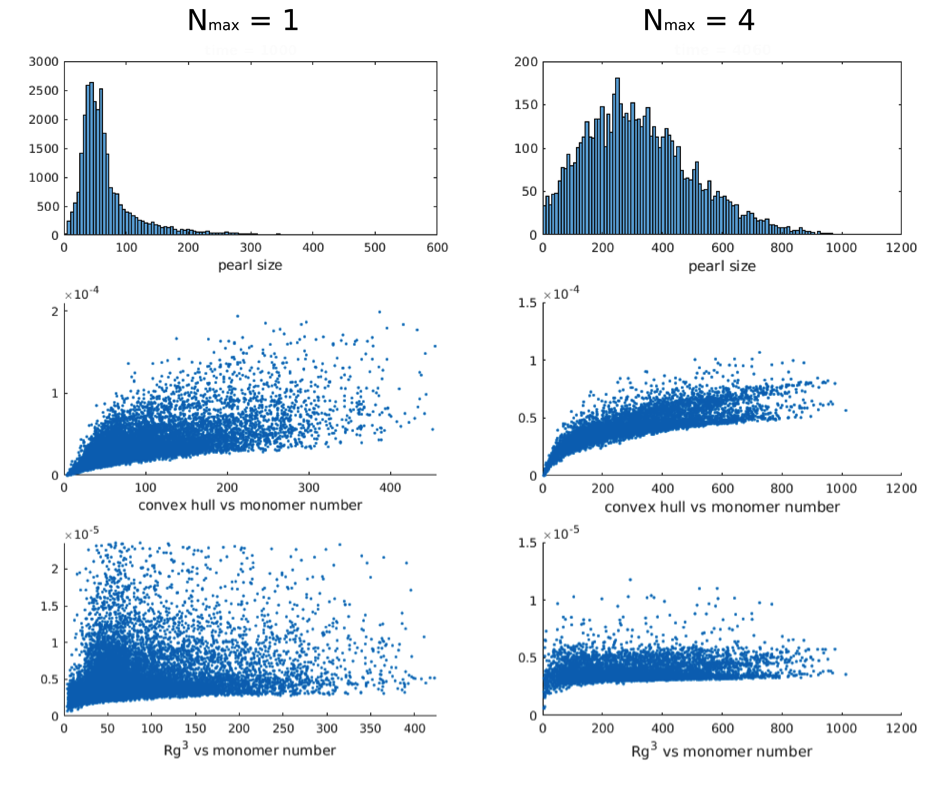}
    \caption{From top to bottom: distribution of pearl sizes, convex
      hull in function of number of monomers in the pearl and radius of
      gyration to the cube in function of number of monomers in the pearl. Left
      panels are for $p = 0.05$ and $N_{max} = 1$, right panels for $p
      = 0.1$ and $N_{max} = 4$}
    \label{fig:louv}
\end{figure}

Pearls were
detected automatically using the Louvain clustering algorithm~\citeltex{blondel2008fast} on
the contact map of the crosslinked polymer. For two markedly different values of the
crosslink probability $p$, we found that the number of monomers per
pearl displays a peaked distribution with a mean corresponding closely
to the measured value of the pearling length $\xi(p)$ (top panels on
the Suppl. \figref{fig:louv}). 
We computed the volume of the pearls, either using the radius of
gyration $R_g$ or the convex hull surrounding each pearl.
We found that the volume of the pearls increases
with the number on monomers found in each pearl (middle and bottom
panels).

\subsection{Short-scale exponent $\gamma_{lim}$ at the end of fixation ($t \gg t^*$)}

\begin{figure}[h]
    \centering
    \includegraphics[width=0.85\linewidth]{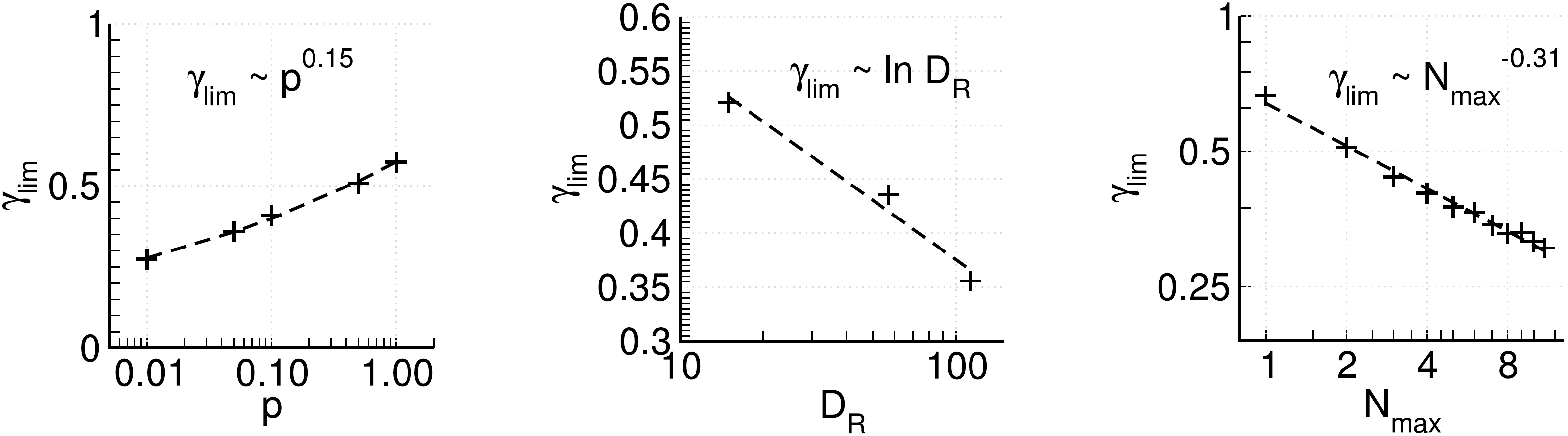}
    \caption{Dependence of $\gamma_{lim}$ on $p$, $D_R$ and
      $N_{max}$.}
    \label{fig:glim}
\end{figure}

The measured $\gamma_0$ after the  pearling transition ($t \gg t^*$) is not a
constant, but depends on different simulation parameters. For large
values of $N_{max}$, $p$ and $D_R$, $\gamma_0$ dynamically and transiently
goes toward 0, before plateauing at final values $\gamma_{lim}$ (see
figure~3F, 
% MANUALREF
%
main text). Suppl. Figure~\ref{fig:glim} highlights the dependence of
this value on the relevant dynamical parameters, suggesting a
characteristic final phase with fractal properties.

\subsection{Effects of additional steric interactions}

We repeated some
simulations for a few sets of parameters (only a few due to heavy
computational requirements) taking into account excluded
volume. The results, presented below, show that despite the fact that
the initial and final values for $\gamma_0$ are (as expected)
different with excluded-volume, the cooperative effect and the scaling
of the dynamics are not significantly modified.
Specifically, we included a
hard-sphere potential between monomers (beads) whose range matches
the average distance between consecutive monomers. This leads to a
change in the exponent of the equilibrium contact probability curve
$P_0(s)$ from -1.5 (random walk) to -2.2 (four-legged polymer-loop
exponent,~\citetltex{marenduzzo2006entropy}). While the initial state of
the polymer
is different, its collapse follows a  dynamics very similar to the
phantom-case dynamics described in our manuscript. Interestingly, the
final state inside the pearls exhibits an exponent $\gamma_0 = 1$,
that could correspond to a crumple globule (also called fractal
globule) as originally described by~\citetltex{grosberg1988role}. 

\begin{figure}[h]
    \centering
    \includegraphics[width=0.85\linewidth]{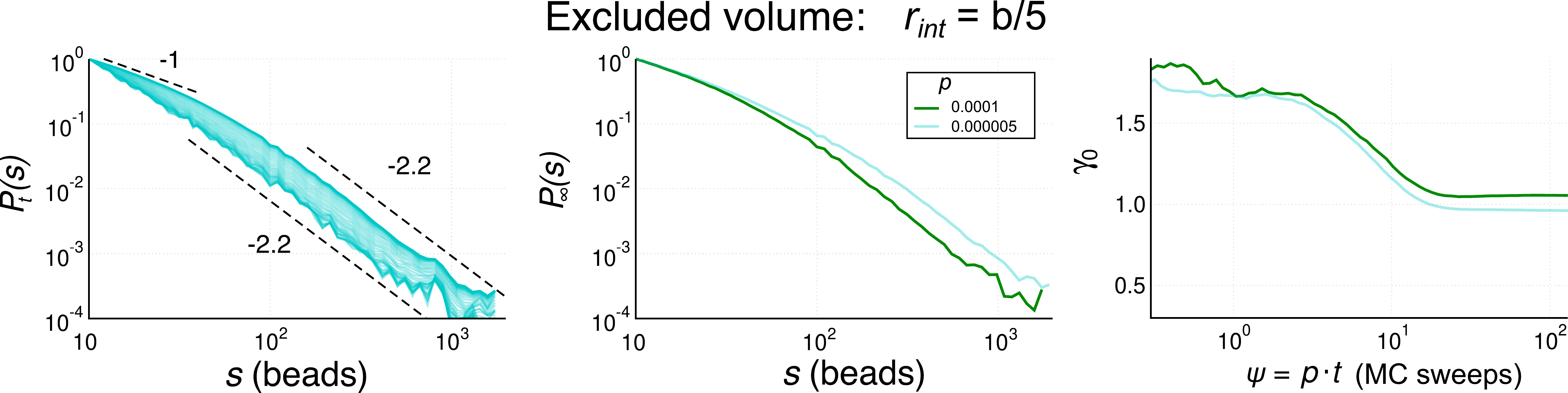}
    \caption{Result of adding hard-core repulsive interactions, from
      left to right: 1) the evolution of $P_t(s)$ in function of
      crosslinking time with $p = 0.0001$, 2) the adsorbing $P_\infty
      (s)$ at different values of $p$ and 3) the dynamics of $\gamma_0$
      in function of $t$ at different values of $p$.}
    \label{fig:volexcl}
\end{figure}

\subsection{Videos of the simulations}

Videos of the simulations are available at the following link:

\url{https://github.com/scovit/crosslink/tree/master/videos}\\
The screen of the video is divided in 4 quadrants, from left to right
and top to bottom. The 
first quadrant show the evolution of the HiC matrix for the single
polymer in time; the second quadrant is the video of the polymer
simulation where each bead is in red, the third and fourth quadrant
show the polymer 
at the bead level (in grey) superimposed with beads calculated by the
center of mass of blocks of respectively 8 and 126 beads, to highlight
the different relaxation dynamics at different scales. The video are
filmed at real computing time, with parameters $p = 1.0$ and $N_{max}
= 1$.

\subsection{Analytical phenomenology of a scale-free Rouse model}
\label{sec:poldyn2}

\subsubsection{The Rouse equation}

The Newton equation  of motion with viscous friction and random noise for a point
particle standing at position $\vec R(t)$ at time $t$ is:
\begin{equation}
  \begin{split}
    F^i(t) &= m \frac{\partial^2 R^i}{\partial t^2}
    - \zeta \frac{\partial R^i}{\partial t} + \eta^i(t),\\
    \mathrm{with}\ \langle \eta^i_n(t) \rangle = 0 & \ \ \
    \mathrm{and} \ \ \ \langle \eta^i_n(t) \eta^j_{n'}(t')
    \rangle = 2\zeta k_B T \delta^{ij} \delta_{nn'} \delta(t - t')
  \end{split}
\end{equation}
In overdamped conditions we can drop the inertial term
$m \frac{\partial^2 R^i}{\partial t^2} \simeq 0$. For a system of many
particles, in the absence of external forces and when the point particle (a bead) belongs to a
Gaussian-potential chain (linking the bead indexed by $n$ with $n-1$
and $n+1$) of Young modulus $K$, we get the stochastic equation for
the free discrete polymer, also called  the Rouse equation:
\begin{equation}
\begin{split}
 \zeta \frac{\partial R^i_n(t)}{\partial t} =
  K \left[R^i_{n+1}(t) + R^i_{n-1}(t) - 2 R^i_n(t)\right] +
  \eta^i_n(t),\\ \mathrm{with}\ \langle \eta^i_n(t) \rangle = 0 \
  \mathrm{and} \ \langle \eta^i_n(t) \eta^j_{n'}(t')
  \rangle = 2\zeta k_B T \delta^{ij} \delta_{nn'} \delta(t - t')
\end{split}
\label{eq:langediscr}
\end{equation}

\subsubsection{Scaling laws and relaxation times}
\label{seq:relaxexac}
%
% other approaches which could be
% investigated include the Fokker-Planck (particularly interesting
% because it works on distributions and densities) and the Path-integral
% (which allows for perturbation or renormalization group) approaches.
%
% For each component of the position of the beads of the polymer, the
% Langevin equation is: 
% $$
% \zeta \frac{\partial R^i_n(t)}{\partial t} =
%  K \left[R^i_{n+1}(t) + R^i_{n-1}(t) - 2 R^i_n(t)\right] +
%  \eta^i_n(t),
% $$
%
%We model the discrete stochastic equation (Eq. \ref{eq:langediscr}) for
Considering the polymer chain as a
continuous and infinite succession of
segments of infinitesimal length fluctuating in the three-dimensional
space in over-damped conditions (no inertial contributions) under the
influence of random forces, the discrete stochastic \eqref{eq:langediscr}  can be approximated by:
\begin{equation}
\zeta \frac{\partial R^i(n, t)}{\partial t} =
 K \frac{\partial ^2 R^i(n,t)}{\partial n^2} + \eta^i(n,t),
% \quad
%\mathrm{and}
%\quad \left. \frac{\partial R^i(n,t)}{\partial n} \right|_{n = 0,N}
%= 0,
\label{eq:contlang}
\end{equation}
where $R^i(n, t)$ is the $i$-th component of the position corresponding
 the $n$-th internal  degree of freedom at time $t$;  $\zeta$ and $K$ are
parameters (possibly  temperature-dependent) with dimension of
friction and energy, respectively,  and not necessarily equal to the
solvent parameters as measured by bead probes.

We can use the Fourier transformation, basically passing to the conjugate variable
 $p$ through the identities:
\begin{equation}
\tilde R^i(p,t) = \frac{1}{\sqrt{2 \pi}L}\int_{\mathbb{R}} \! \mathrm{d} n \,
e^{-i 2 \pi p n} \,  R^i(n,t), \ \ \mathrm{and} \ \ 
R^i(n, t) = \frac{L}{\sqrt{2 \pi}}\int_{\mathbb{R}} \! \mathrm{d} p \,
e^{i 2 \pi p n} \,  \tilde R^i(p, t)
\label{eq:fourier}
\end{equation}
%the integral, as well as the whole model, is a formal tools to make
%algebraic calculation, the domain is not defined,
$L$ represents the value of the
Dirac delta in zero: $\left. \delta(p) \right|_{p=0} = L$
and  $\left. \delta(n) \right|_{n=0} = L^{-1}$. It can be formally
interpreted through a limit-integral representation of the delta
function in term of box functions, and it can in fact be chosen
arbitrarily (the discretization of the polymer for instance or a unit
of measurement make sense). Each component $\tilde R^i(p, t)$, defined
for $p \ne 0$ is
related to wavelength of size $1/p$. The following relations relate
the Fourier transform of unit functions and the Dirac delta:
$$
\int_{\mathbb{R}} {\rm
  d}p\,  e^{i 2 \pi p  n} = \delta (n),\ \ {\rm and}\ \ 
\int_{\mathbb{R}} {\rm d}n\,  e^{-i 2 \pi p  n} = 
\delta(p).
$$
% the evolution of $\tilde  R^i(p, t)$ in time univocally determine an
% instance of the evolution of the conformation $R^i(n, t)$ in the
% deterministic equation corresponding to eq. \ref{eq:contlang}.

Under those assumptions, $\eta^i(n, t)$ in \eqref{eq:contlang} is a
random force that satisfies the following two 
conditions:  (i) the process $\eta^i(n, t)$ is a Gaussian process and
 (ii) it is Markovian, namely  its correlation time is
infinitely short:
% (\rev{cite Kubo and his citations}):
$$
\langle \eta^i(n, t) \eta^j(n', t') \rangle = 4 \pi G_\eta \,\,
   \delta^{ij}\, L \delta(n - n') \delta(t-t'),
$$
where $G_\eta$ is a constant with dimension of a force multiplied
by an impulse,
and the mean $\langle \cdot \rangle$ is taken over an ensemble of
polymers.

Applying the transformation we turn \eqref{eq:contlang} into the
following linear stochastic equation:
\begin{equation}
\zeta \frac{\partial \tilde R^i(p,t)}{\partial t} = 
  - 4 \pi^2 p^2 K \tilde R^i(p,t) + \tilde \eta^i(p,t),\ 
\label{eq:genlang}
\end{equation}
where modes $p$ are decoupled. The Fourier transform of the random force has
the following property:
\begin{equation}
\langle \tilde \eta^i(p, t) \tilde \eta^j(q, t')^* \rangle =
 2 G_\eta \,\, \delta^{ij} \, \frac{1}{L}
\delta(p - q) \delta(t - t'),
\label{eq:white}
\end{equation}
which makes \eqref{eq:genlang} a classical Langevin equation. 
The calculation of the values of $G_\eta$ is detailed in the next section
(\ref{seq:therm}).
%then
%we can solve the intensity of the random force using the
%fluctuation-dissipation theorem: $G_\eta = $.

The time correlation function in Fourier space is given by
\begin{equation}
\langle \tilde R^i(p, t_0) \tilde R^j(q, t + t_0)^* \rangle = 
\left\langle \left| \tilde R^i(p, t_0) \right|^2 \right\rangle
e^{- p^2 t/\tau} \, \delta_{ij}\, \frac{1}{L} \delta(p - q),
\quad \mathrm{with} \quad \tau = \frac{\zeta}{4 \pi^2 K},
\label{eq:relaxprec}
\end{equation}
which highlights the hierarchy of polymer  relaxation modes:
from smaller wavelength (faster) to longer wavelength (slower).
%The slower mode, with $p = 1$, have relaxation time equal to the
%Rouse
%time $\tau$.
Also, assuming the energy equipartition law,
\begin{equation}
\left\langle \left| \tilde R^i(p, t_0) \right|^2 \right\rangle =
\frac{k_B T}{4 \pi^2 K} \frac{1}{p^2}.
\label{eq:equipart}
\end{equation}

We can calculate the diffusion of a segment of the polymer:
$$
\left\langle \left| \vec R(0, t_0) - \vec R(0, t + t_0) \right|^2
\right\rangle = 
6 \left\langle \left| R^i(0, t_0) \right|^2 \right\rangle - 
6 \left\langle \left| R^i(0, t_0)R^i(0, t + t_0)  \right|
\right\rangle =
$$
$$
= \frac{3 k_B T L}{4 \pi^3 K} \int_{\mathbb{R}} \!\!{\rm d}p\; \frac{1}{p^2}
- \frac{3 k_B T L}{4 \pi^3 K} \int_{\mathbb{R}} \!\!{\rm d}p\; \frac{1}{p^2} 
e^{- p^2 t/\tau} =
$$
\begin{equation}
=\frac{3 k_B T L}{4 \pi^3 K} \int_{\mathbb{R}} \!\!{\rm d}p\; \frac{1}{p^2} \left(
1 - e^{- p^2 t/\tau} \right)
= \sqrt{\frac{1}{K \pi^3 \zeta}}\, 3k_B T L \cdot t^{1/2}
\end{equation}
by integrating by parts. It corresponds to Rouse sub-diffusion,
whose coefficient in this model is equal to $C_t = 3 k_B T L /
\sqrt{K \pi^3 \zeta}$ and is a measurable quantity.

At fixed time, we can calculate the  point-to-point 
mean square distance:
$$
\left\langle \left| \vec R(s_0, 0) - \vec R(s + s_0, 0) \right|^2
\right\rangle = 6 \left\langle \left| R^i(s_0, 0) \right|^2 \right\rangle - 
6 \left\langle \left| R^i(s_0, 0)R^i(s + s_0, 0)  \right|
\right\rangle =
$$
$$
=\frac{3 k_B T L}{4 \pi^3 K} \int_{\mathbb{R}} \!\!{\rm d}p\; \frac{1}{p^2}
- \frac{3 k_B T L}{4 \pi^3 K} \int_{\mathbb{R}} \!\!{\rm d}p\; \frac{1}{p^2} 
e^{i 2 \pi p s} =
$$
\begin{equation}
= \frac{3 k_B T L}{2 K \pi} \cdot \left| s \right|
\label{eq:secondmom}
\end{equation}
by integrating by parts and through Residue theorem. This equation reproduces
the well known random-walk behavior, whose coefficient $C_s = 3 k_B T
L / (2 K \pi)$ is a measurable quantity. From a micro-rheological
point-of-view, we can determine model parameters by fitting the scaling of
the Rouse sub-diffusion and  the scaling of the point-to-point
mean square distance:
\begin{equation}
\label{eq:viscoela}
\zeta = \frac{6}{\pi^2}\frac{C_s}{C_t^2} L k_B T,
\quad {\rm and} \quad
K = \frac{3}{2 \pi}\frac{1}{C_s} L k_B T,
\quad {\rm implying} \quad
\tau = \frac{1}{\pi^3}\left(\frac{C_s}{C_t}\right)^2;
\end{equation}
Notice that the scaling of the parameters $\zeta$ (and $K$) in terms of
the arbitrary length $L$, emerging from purely algebraic
considerations, is reminiscent of the scaling of viscosity
as a function of  the persistence length of a worm-like-chain $\zeta = 6 \pi
\eta b$ (with $b \propto L$).

We can finally write the relaxation time
\eqref{eq:relaxprec} for each mode $p$
in the following terms:
\begin{equation}
\tau_R(p) = \frac{1}{\pi^3}\left(\frac{C_s}{C_t}\right)^2
\cdot \frac{1}{p^2}
\label{eq:relaxexactp}
\end{equation}
which is a model-independent prediction (zero parameters, only
observable coefficients) of the theory.

\subsubsection{Calculation of the thermal noise intensity $G_{\eta}$}

\label{seq:therm}
It is possible to verify by substitution that the solution of equation
\ref{eq:genlang} is:
$$
\tilde R^i(p,t) = \tilde R^i(p,0) e^{-p^2 \frac{t}{\tau}} +
\frac{1}{\zeta}
\int_0^t {\rm d} t' e^{-p^2 \frac{t - t'}{\tau}} \tilde \eta^i(p, t');
$$
Using this relation we can calculate the mean thermal fluctuation over
a long time:
$$
\overline{\left| \tilde R^i(p,T) \right|^2 } = 
\lim_{T \to \infty} \frac{1}{T} \left(
\left| \tilde R^i(p,0) \right| e^{-2 p^2 \frac{t}{\tau}} +
\frac{1}{\zeta^2}
\int_0^t \int_0^t {\rm d} t' {\rm d} t''
e^{-p^2 \left[ \frac{t - t'}{\tau} + \frac{t - t''}{\tau} \right]}
\tilde \eta^i(p, t') \tilde \eta^i(p, t'')^*
\right),
$$
Finally, assuming for solving the problem at equilibrium:

\noindent
1 - {\it ergodicity}
($\overline{\left| \cdot \right|} \to \langle \cdot \rangle$),

\noindent
2 - the white noise definition \eqref{eq:white}, and

\noindent
3 - the energy equipartition law \eqref{eq:equipart}, we find:
$$
\frac{k_B T}{4\pi^2 K} \frac{1}{p^2} =
\left\langle \left| \tilde R^i(p, t_0) \right|^2 \right\rangle =
\overline{\left| \tilde R^i(p,T) \right|^2 } =
G_\eta \frac{2}{\zeta^2} \lim_{t \to \infty} \int_0^t {\rm d} t' 
e^{- 2 p^2 \frac{t - t'}{\tau}}
$$
$$
\frac{k_B T}{4\pi^2 K} \frac{1}{p^2} =
G_\eta \frac{\tau}{\zeta ^2 p^2}
$$
so that
\begin{equation}
G_\eta = \zeta k_B T.
\end{equation}

\subsubsection{Relaxation times for out-of-equilibrium loops}
\label{sec:relaxexact}

For looped configurations, everything that has been written in the
previous paragraph holds true apart that, focusing at the loop level,
the additional border condition imposes:
\begin{equation}
\tilde R^i(p, t) < \epsilon\qquad \mathrm{for\ all}\qquad
 p \ne k \cdot \frac{1}{2s}
\end{equation}
with $\epsilon$ small, $k \in \mathbb Z_{> 0}$, and $s$ the loop
arc-length.
The surviving modes follow the same relaxation dynamics of 
\eqref{eq:relaxprec} and \eqref{eq:relaxexactp}. As such, the longer
relaxation time for the whole loop corresponds to $p = 1 / (2s)$. We
conclude that the scaling of the relaxation times for the loops as a
function of arc-length is:

\begin{equation}
\tau_R(s) = D_R^{-1} \cdot s^2 =
\frac{4}{\pi^3}\left(\frac{C_s}{C_t}\right)^2 \cdot s^2,
\label{eq:relaxexact}
\end{equation}
as verified in  our simulations through the scaling of $\xi$ in $t^*$.

\let\oldaddcontentsline\addcontentsline% Store \addcontentsline
\renewcommand{\addcontentsline}[3]{}% Make \addcontentsline a no-op
\bibliographystyleltex{aipauth4-1}
\bibliographyltex{biblio}

\end{document}